\documentclass{aa}
\usepackage {epsfig,graphicx,color}
\usepackage {txfonts  }
\usepackage {natbib   }
\usepackage {float    }
\usepackage {enumerate}
\usepackage {tabularx }

\begin{document}

%

\title {3D simulations of globules and pillars formation around HII regions: turbulence and shock curvature
}


\author{ 
P. Tremblin  \inst{1}\and 
E. Audit     \inst{1,2}\and 
V. Minier    \inst{1}\and   
W. Schmidt   \inst{3}\and
N. Schneider \inst{1}      
       }

\institute{Laboratoire AIM Paris-Saclay (CEA/Irfu - Uni. Paris Diderot - CNRS/INSU), 
           Centre d'\'etudes de Saclay,  91191 Gif-Sur-Yvette, France    
           \and
           Maison de la Simulation, CEA-CNRS-INRIA-UPS-UVSQ, USR 3441,
           Centre d'\'etude de Saclay, 91191 Gif-Sur-Yvette, France
           \and
           Institut f\"ur Astrophysik der Universit\"at G\"ottingen,
           Friedrich-Hund-Platz 1, D-37077 G\"ottingen, Germany
           }

\date{
}
\mail{pascal.tremblin@cea.fr}

\titlerunning{Globules and Pillars Simulations}
\authorrunning{Tremblin et al.}

\abstract{}
{We investigate the interplay between the ionization radiation from
  massive stars and the turbulence inside the surrounding molecular
  gas thanks to 3D numerical simulations.}
{We used the 3D hydrodynamical code HERACLES to model an initial
  turbulent medium that is ionized and heated by an ionizing
  source. Three different simulations are performed with different
  mean Mach numbers (1, 2 and 4). A non-equilibrium model for the
  ionization and the associated thermal processes was used. This
  revealed to be crucial when turbulent ram pressure is of the same
  order as the ionized-gas pressure.}
{The density structures initiated by the turbulence cause local
  curvatures of the dense shell formed by the ionization
  compression. When the curvature of the shell is sufficient, the
  shell collapse on itself to form a pillar while a smaller curvature
  leads to the formation of dense clumps that are accelerated with the
  shell and therefore remain in the shell during the simulation. When
  the turbulent ram pressure of the cold gas is sufficient to balance
  the ionized-gas pressure, some dense-gas bubbles have enough kinetic
  energy to penetrate inside the ionized medium, forming cometary
  globules. This suggests a direct relation in the observations
  between the presence of globules and the relative importance of the
  turbulence compared to the ionized-gas pressure. The probability
  density functions present a double peak structure when the
  turbulence is low relative to the ionized-gas pressure. This could
  be used in observations as an indication of the turbulence inside
  molecular clouds.}
{}

\keywords{Stars: formation - HII regions - ISM: structure - Turbulence - Methods: numerical}

\maketitle


%
%

\section{Introduction}

Although massive stars have a great impact on their environment, the
importance of their radiative feedback on star-formation rates is
still a matter of discussion. For example, \citet{Dale:2005fa} found a
slight enhancement of the star-formation activity when the radiative
feedback is taken into account whereas they recently presented
simulations on a scale of a giant molecular cloud with almost no
impact \citep[see][]{Dale:2011ct}. A closer look at the small scales
helps addressing this question by identifying the detailed mechanisms
that form the dense structures eventually leading to star
formation. These structures consist mainly of pillars of gas pointing
toward the ionizing source, globules detached from the parent
molecular cloud, and dense clumps at the interface between the HII
region and the cloud. Optical, infrared (IR), and far-IR observations,
using the Hubble space telescope and the Spitzer and Herschel
satellites impressively revealed these features, mainly in high-mass
star-forming regions
\citep[e.g.][]{Hester:1996ir,Gerin:2009ib,Deharveng:2010cp,Schneider:2010ec,Zavagno:2010jv}. Several
models have been proposed to explain their formation. The collect and
collapse scenario described by \citet{Elmegreen:1977iq} or shadowing
instabilities in the ionization front \citep[e.g.][]{Williams:1999bv}
concentrate on the formation of dense clumps at the
interface. \citet{Bertoldi:1989bq} and \citet{Lefloch:1994ts} with the
radiation driven implosion of clumps studied the formation of globules
and \citet{Mackey:2010cv} proposed the shadowing effect to explain the
formation of pillars.\\
\indent In a previous paper (\citet{Tremblin:2012ej}, paper I
hereafter), we presented a new approach emphasizing the importance of
the curvature of the dense shell at the edge of the HII region to
explain the formation of pillars and dense clumps in the shell. When
the shell is sufficiently curved, a pillar will form by the collapse
of the shell on itself, while a smaller curvature will trigger lateral
flows forming dense clumps and dips inside the shell. However this
scenario and the previous ones presented above are rather idealized
set-ups and have to be validated in more realistic situations,
e.g. taking into account a turbulent molecular cloud. \\ 
\indent Recent studies
\citep[see][]{Mellema:2006if,Gritschneder:2010du,Arthur:2011ck}
started to concentrate on the interplay between ionization from the
massive stars and the turbulence inside the molecular cloud. They
found that pillars, dense clumps and globules emerge naturally in
their models, however the detailed processes forming these structures
are difficult to identify with the turbulence.\\ 
\indent In the present study, we concentrate on the interplay between
turbulence and ionization in the light of the detailed processes
presented in paper I. We first present the numerical method used and
some tests of the out-of-equilibrium ionization module,
followed by three ionized turbulent simulations at different mean Mach
numbers (1, 2 and 4) and finally the comparison with the previous
idealized situations presented in paper I and a new approach to
explain the formation of globules. Finally we show that probability
density functions of the gas can be used in observations to determine
the importance of turbulence relative to the ionized-gas pressure.

%
%

\section{Numerical methods}\label{num_met}
We used the HERACLES
code\footnote{http://irfu.cea.fr/Projets/Site\_heracles/index.hmtl}
\citep{Gonzalez:2007gd} with the ionization module described in paper
I and the cooling module described in \citet{Audit:2010jx}
  (based on \citet{Wolfire:2003hf}) to model the molecular cloud and
the ionization coming from the OB cluster. In addition to the physics
of paper I, the turbulence in the cloud is modeled using the
turbulence module described in \citet{Schmidt:2006gf} and
\citet{Schmidt:2009dr}. Gravity is an important ingredient for
  ultra-compact HII region \citep[e.g.][]{Peters:2010bo}, however we
  are interested in large scale HII regions (e.g. the Rosette Molecular
  cloud, \citet{Schneider:2010ec}. Self-gravity and the gravity from
  the ionizing source were considered in paper I without noticeable
  change on the formation of the structures. Therefore we do not
  consider gravity in the present study.\\ 
\begin{figure}[h]
\centering
\includegraphics[width=\linewidth]{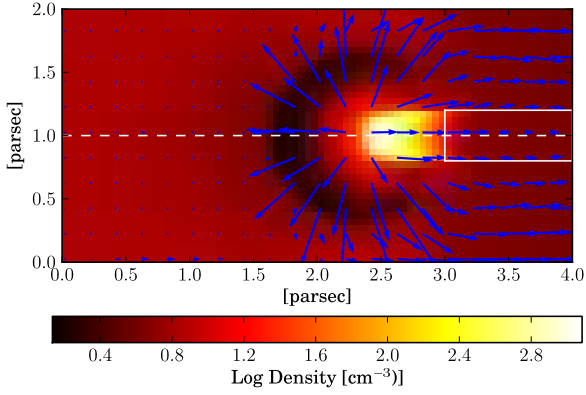}
\caption{\label{tests} Cut of the density field and velocity field in the
  radiation driven implosion of a clump inside an ionized medium. The
  gas behind the clump in the white box is out of equilibrium and
  recombines slowly. The gas at the front of the clump is escaping by
  the rocket effect and is cooled down by the strong expansion.} 
\end{figure}
\begin{figure}[!h]
\centering
\includegraphics[width=\linewidth]{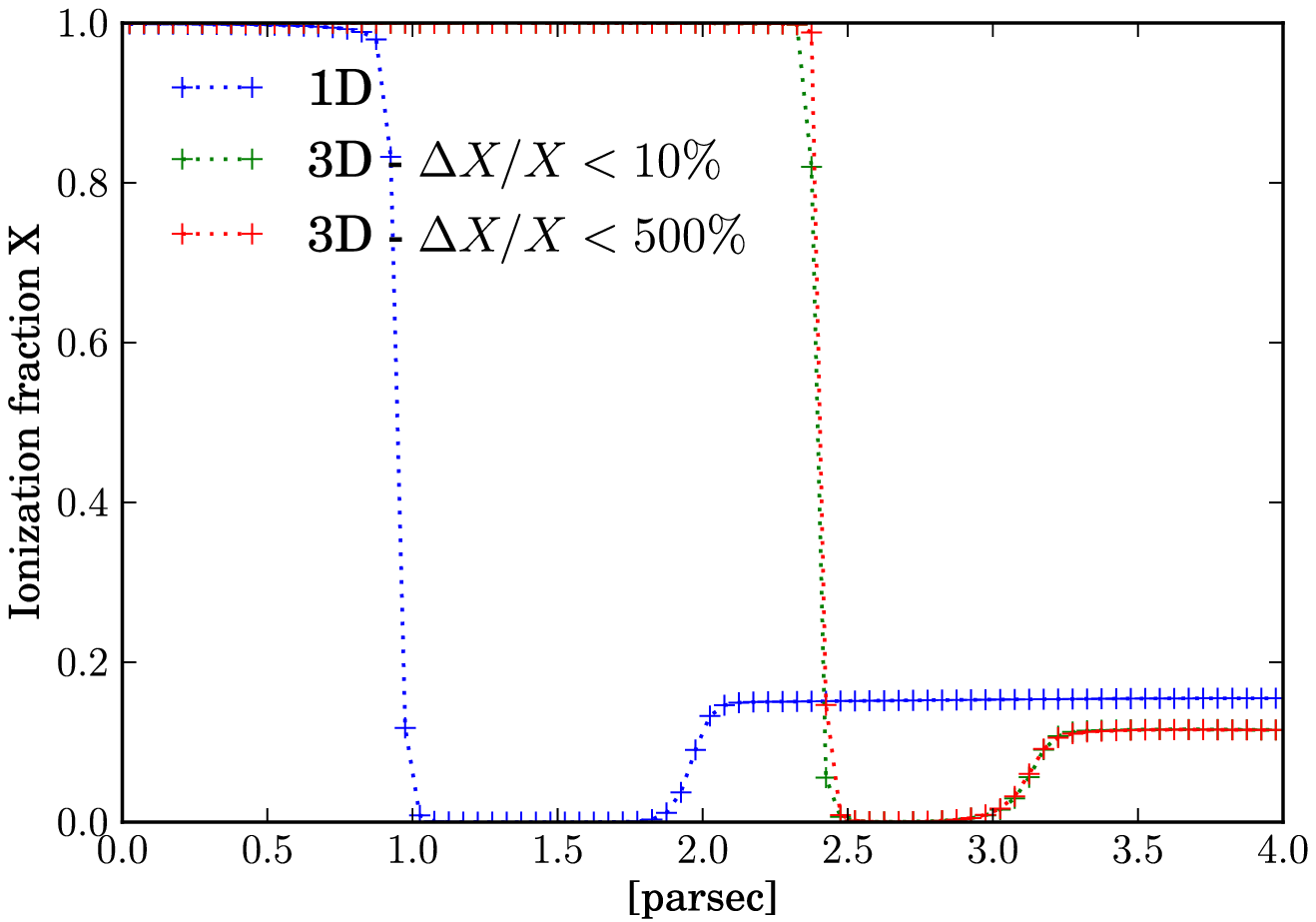}
\includegraphics[width=\linewidth]{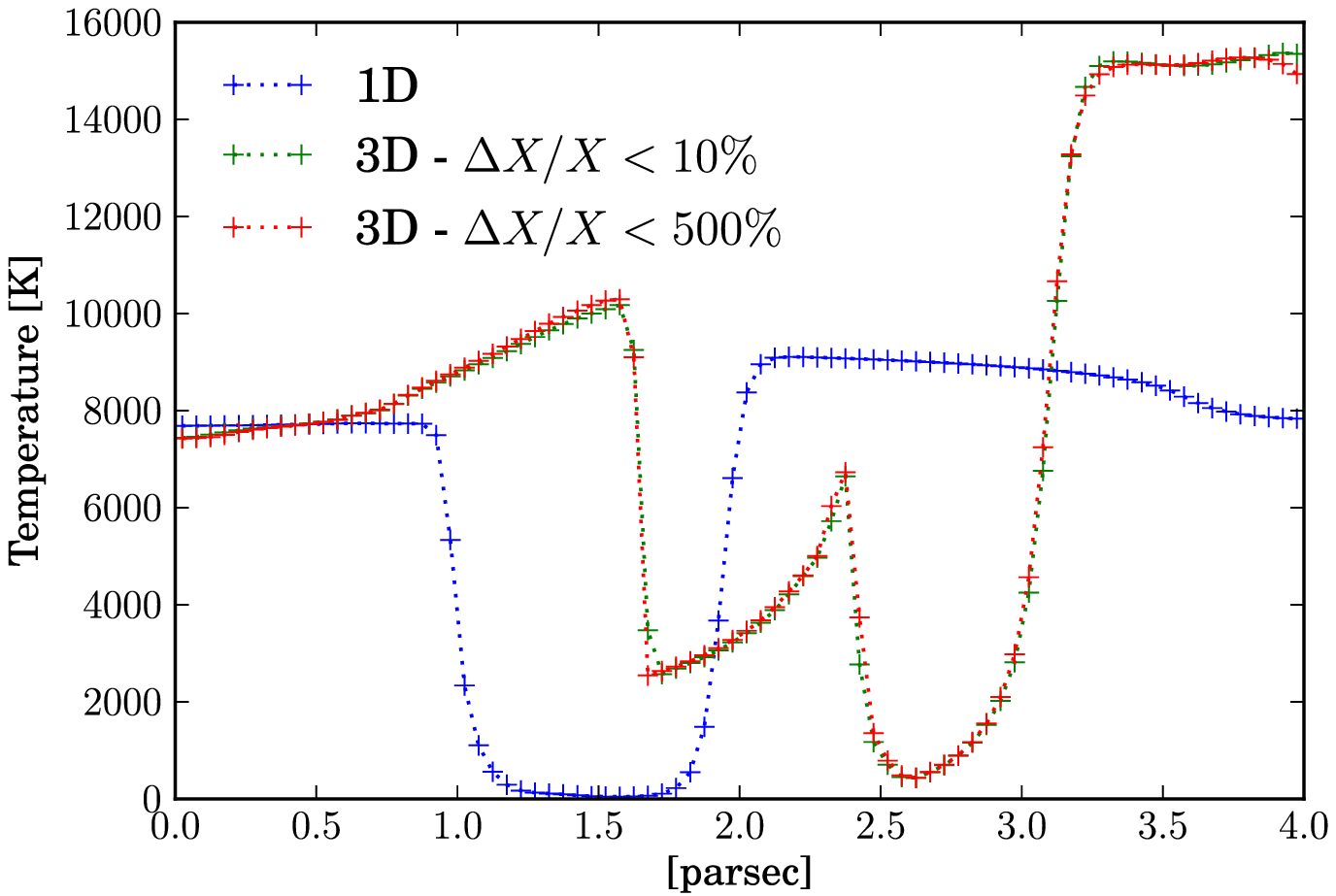}
\caption{\label{algo_XT} 1D and 3D simulations of an isolated clump
  inside a completely ionized medium exposed to an ionization flux of
  10$^9$ photons per second per squared centimeter. Note that in the
  1D simulation the clump is in fact an infinite slab. Top: ionization
  profile through the middle of the box at t = 500 ky. Bottom
  Temperature profile at T = 500ky. The ionization source is at x = 0
  parsec. In the 1D simulation the clump is located at x = 1.5 parsec
  while in the 3D simulations it is at 2.75 parsec. The two 3D
  simulations show that it is not needed to limit too much the
  variation of the ionization fraction. The gas behind the clump is in
  the shadow and is therefore being recombined. Its recombination time
  is long and is not at equilibrium. These simulations also show the
  importance of the 3D effects on the temperature structure.}
\end{figure}
\indent First, we describe the model we used for ionization. It is an
out-of-equilibrium model solving the equations
\begin{eqnarray}\label{io_eq}
d (F_\gamma)/dx &=& -n_H(1-X)\sigma_\gamma F_\gamma + \delta_0(x) S_*\cr
d (Xn_H)/dt &=& I - R = n_H(1-X)\sigma_\gamma F_\gamma - \beta X^2n^2_H\cr
de/dt &=& I\times e_\gamma - R\times k_bT/(\gamma-1)
\end{eqnarray}
in which $X$ is the ionization fraction (advected with the
hydrodynamic), $F_\gamma$ is the photon flux, $S_*$ the source term,
$n_H$ the total hydrogen density, $\sigma_\gamma$ the cross-section
for ionization, $I$ the ionization rate, $R$ the recombination rate,
$\beta$ is given by $2\times 10^{-10}T^{-0.75}$cm$^3$/s, $e$ is the
internal energy of the gas, $e_\gamma$ the energy transferred from the
ionizing photons to the electrons and $T$ the temperature of thermal
equilibrium between all the species. The first equation is the ray
tracing of the ionizing photons coming from the source. The second one
is the ionization/recombination photo-chemistry evolution and the
third one is the thermodynamic evolution associated. 1D and 3D
  tests of the out-of-equilibrium ionization and recombination
  processes are performed in Sect. \ref{tests_sect}.\\ 
\indent The turbulent forcing is modeled by an Ornstein-Uhlenbeck process in Fourier
space. For each mode of the force field, random increments with a
Gaussian distribution are added after each time step. The wave-numbers of 
 the forcing modes are in the interval $[0,2k_0]$, where $k_0$ is given 
by the typical length scale of the forcing, $L=2\pi/k_0$ with $L=4$
pc, the size of the box in our case.  
By projecting the modes in transversal and longitudinal 
directions with regard to the wave vectors, divergence-free and 
rotation-free components are produced in physical space. The resulting 
force field is statistically homogeneous, isotropic, and stationary. The 
RMS magnitude is of the order $V^2/L$, where $V$ is the characteristic 
velocity of the flow. We will perform three simulations with $V$ resulting
in stationary transonic and supersonic velocity: Mach 1, 2 and 4. The
turbulence used is solenoidal with a ratio of compressive forcing
power to the total forcing power of 10 \%. \\ 
\indent Three turbulent scenarii are investigated in Sect.
\ref{turbulence}, a transonic turbulent molecular cloud (mean Mach
number around 1) and two supersonic cases (mean Mach number of 2 and
4). The box is a cube of 4 parsec at a resolution of 0.01 pc
  (400$^3$), the mean density is 500 cm$^{-3}$ and the
  temperature is initially at 25 Kelvin. During the turbulent
  evolution, the boundary conditions are periodic. 
The simulations are ran until a statistically steady state is
obtained, i.e. a constant mean Mach number in the box. Then we turn on
the ionization with a flux of 10$^9$ photons per second per squared centimeter, which
is the typical flux of an O4 star at 30 parsecs. The ionization is
coming from the top of the box and this face is changed to reflexive
boundary condition for the hydrodynamic while the opposite side is set
to free flow. The turbulent stirring is kept during the ionization
phase. Some runs were done without the turbulence maintained without
noticeable change, the large-scale modes do not have time to influence
the small scales during the ionization phase. In Sect. \ref{turbulence}, we
investigate the interplay between turbulence and ionization for the
three different simulations.

\section{Tests of the out-of-equilibrium ionization and recombination
  processes}\label{tests_sect} 

When turbulence is included in an ionization simulation, the main
difference is the possibility that some ionized gas gets into the
shadow of the dense unionized gas because of the mixing. This is
typically an out-of-equilibrium state for the ionized gas that will
begin to recombine. To study this state, we investigate
  simplified set-ups in which hot ionized gas is put in the shadow of
  dense cold gas and therefore will be out of equilibrium. We
simulate the impact of ionization on a clump (D$_{clump}$=0.5 parsec)
inside an homogeneous ionized medium (see Fig. \ref{tests}). We also
perform an ''equivalent'' 1D simulation where the clump will be an
infinite slab, to identify the effect of the geometry. The gas in the
shadow of the clump/slab will start to recombine since it is out of
equilibrium. The typical recombination time for the ionized gas is
given by $1/\beta n$ that is of the order of 10$^4$-10$^5$ years for
ionized gas at a density of 1-10/cm$^3$. It is of the order of the gas
dynamical time: D$_{clump}$/c$_{io}$ = 10$^5$ years, in which c$_{io}$
is the sound speed of the ionized gas. Therefore an out-of-equilibrium
model is needed to treat this state of the gas. We studied first the
profile of the ionization fraction and the effect of the
limited-variation time-stepping and then the structure of the
temperature in these tests.\\ 
\begin{figure*}[!t]
\centering
\rotatebox{90}{Initial average Mach number:
  1}\includegraphics[trim=5cm 0.5cm 3cm
  0.5cm,clip,width=\linewidth]{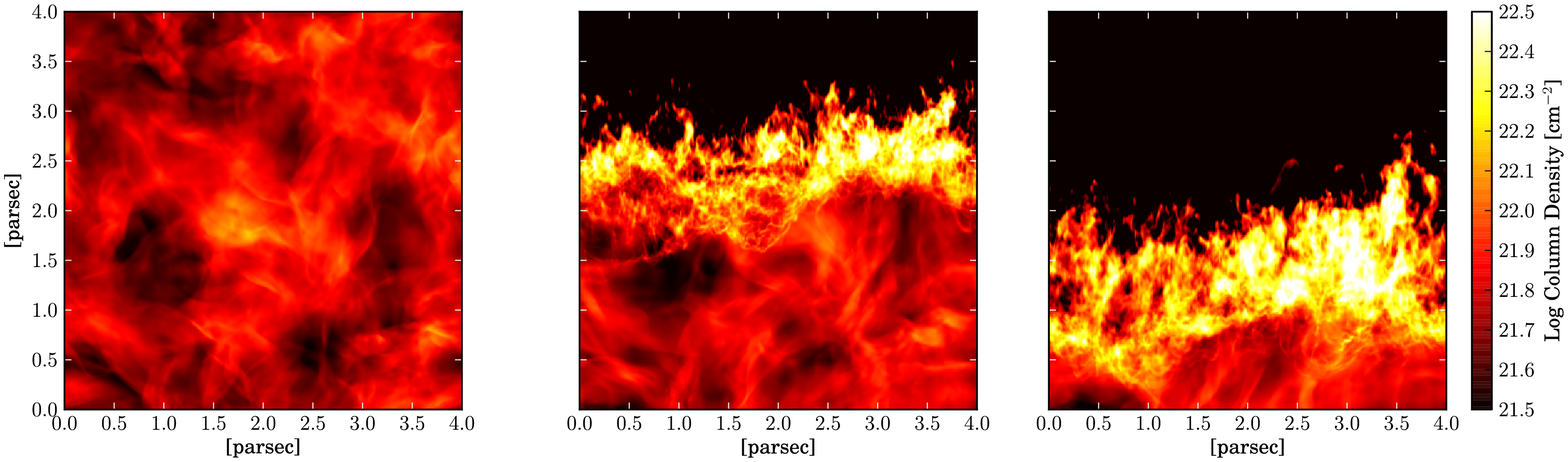} 
\rotatebox{90}{Initial average Mach number:
  2}\includegraphics[trim=5cm 0.5cm 3cm
  0.5cm,clip,width=\linewidth]{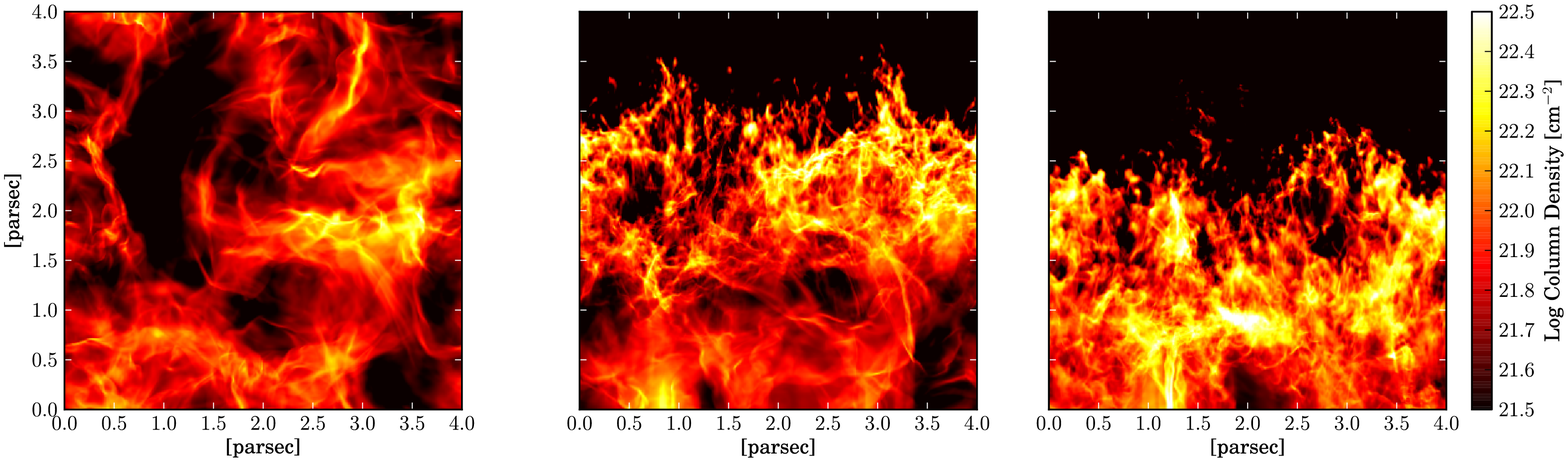} 
\rotatebox{90}{Initial average Mach number:
  4}\includegraphics[trim=5cm 0.0cm 3cm
  0.5cm,clip,width=\linewidth]{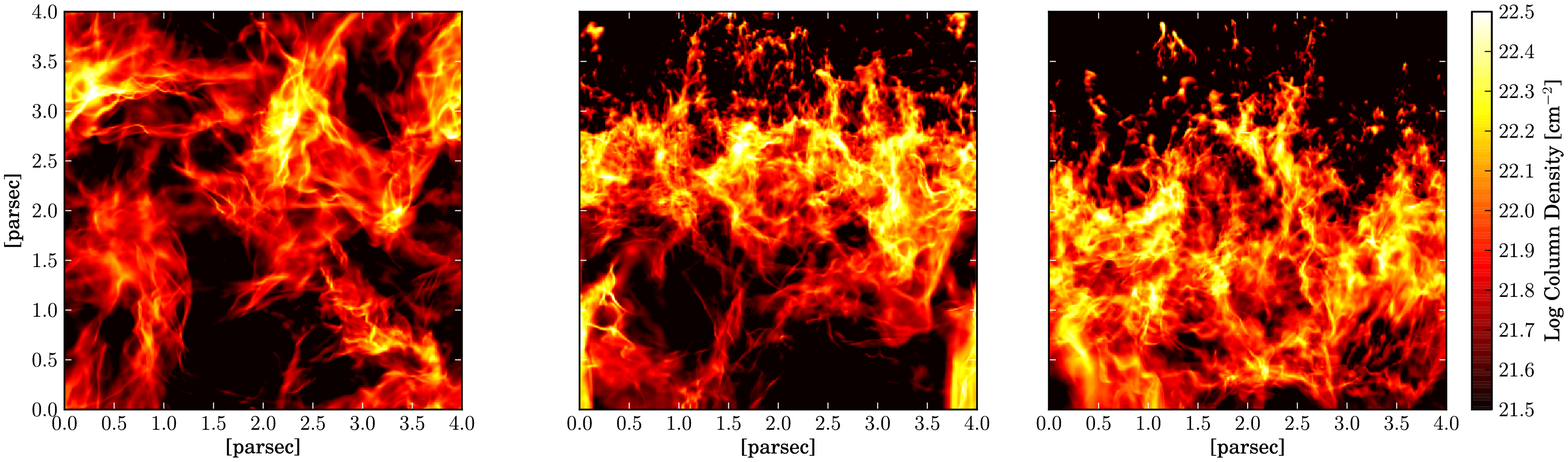} 
\begin{minipage}[c]{0.32\linewidth}
\centering
Time: 0 My
\end{minipage}
\begin{minipage}[c]{0.32\linewidth}
\centering
Time: 0.5 My
\end{minipage}
\begin{minipage}[c]{0.32\linewidth}
\centering
Time: 1 My
\end{minipage}
\caption{\label{crho} Column density snapshots of the three turbulent
  simulations (row 1: Mach 1, row 2: Mach 2, row 3: Mach 4) et three
  different time (column 1: before ionization starts, column 2: 500 ky
  after ionization is turned on, column 3: 1 My). } 
\end{figure*}
\indent Our method solves the two first equations in (\ref{io_eq})
implicitly while the thermal balance is solved explicitly and is
sub-cycled to limit the variation of $T$. Usually the variation of $X$
is limited at 10 \% in the implicit step. However, it is often not
needed to limit that much the variations. The ionization-fraction and
the temperature profiles are plotted in Fig. \ref{algo_XT}, after 500
ky of evolution and for limited variations of 10 and 500 \% and there
is almost no difference between the two. At the front, the equilibrium
for ionization is reached even with long time-step thanks to the
implicit method, the ionization fraction is nearly at one. At the
back, the recombination time of a hot ionized plasma is quite
long. For example in our case the recombination time is of the order
of 100 ky which also explains why the ionization fraction is still
around 0.2 after 500 ky. Therefore it is possible to allow long
time-steps.\\ 
\indent We now investigate the temperature structure. The 3D profiles
in Fig. \ref{algo_XT} are taken through the middle of the box (the
white-dashed line in Fig \ref{tests}), therefore through the middle of
the clump, which is located between 2.5 and 3 parsecs. In the 1D case,
the clump is a slab located after 500 ky between 1 and 2 parsecs. The
difference of position between the two is easy to understand: in the
1D case, the ionized material cannot be evacuated on the side, the
column density in front of the ionization front is higher. There are only a few
ionization processes occurring at the ionization front, the dynamics is dominated by the expansion of the gas evaporating
   from the slab. In the 3D
case, the ionized gas is escaping on the side of the clumps, therefore
ionization can penetrate further inside the clump and deposit more
energy. As a consequence, the clump is pushed further away by rocket
motion and the energy deposit at the surface of the clump leads to a
high expansion of the ionized gas and to a shock surface that can be
seen in Fig. \ref{tests}. The ionized gas is cooled down to 3000-6000
Kelvin by the expansion at the front.\\ 
\indent The temperature structure is also quite different behind the
clump (white box in Fig. \ref{tests}) because of the shadowed
recombination. In the 1D simulation, the thermal equilibrium is
reached at the front, in the HII region.Behind the slab, the
  gas is not at equilibrium and recombines, the ionization fraction is
  decreasing from 1 to 0.2. The pressure drops by nearly a factor of 2 when the ionization
 fraction decreases from 1 to $\approx$ 0.2. Moreover, most of the energy
 emitted during the recombination process is radiated away by the
 cooling, so that the temperature of the gas does not vary much,
 increasing only slightly, from $\approx$ 8000 K to $\approx$ 9000 K.
 The temperature structure is quite different in 3D as shown in
Fig. \ref{algo_XT}. The cooling by recombination behind the clump
should decrease the pressure of the gas as in the 1D
  case. However in 3D, it leads to lateral gas flows from the hot
ionized gas exposed to the radiation to the hot recombined gas in the
shadow. Therefore the recombination behind the clump is done more or
less at constant pressure leading to an increase in temperature behind
the clump, which is surprising at first sight. \\ 
\indent This simple test shows that the equilibrium assumption for
ionization is invalid when hot ionized gas gets in the shadow of dense
unionized gas (Ionization fraction of 20 \% instead of
  0). The equilibrium assumption is usually justified by the fact
that the recombination time is much shorter than the dynamical
time. This is usually true in the HII region, however this is not the
case at the interface between the HII region and the cold gas when
turbulence is included. Indeed especially at high level of
  turbulence, hot ionized gas will be mixed with cold dense gas and
  therefore will be in the shadow of the cold gas, exactly in the
  situation we studied in this simplified test. The ionization
equilibrium is not reached for a non-negligible part of the ionized
gas that can get in the shadow of dense unionized gas. This is
especially the case when the level of turbulence is sufficient to
balance the ionized-gas pressure and to mix the unionized and ionized
phases as we will see in the next section.\\

%
%

\section{Turbulent simulation}\label{turbulence}

\subsection{Transonic turbulence}

\begin{figure}[!h]
\centering
\includegraphics[width=\linewidth]{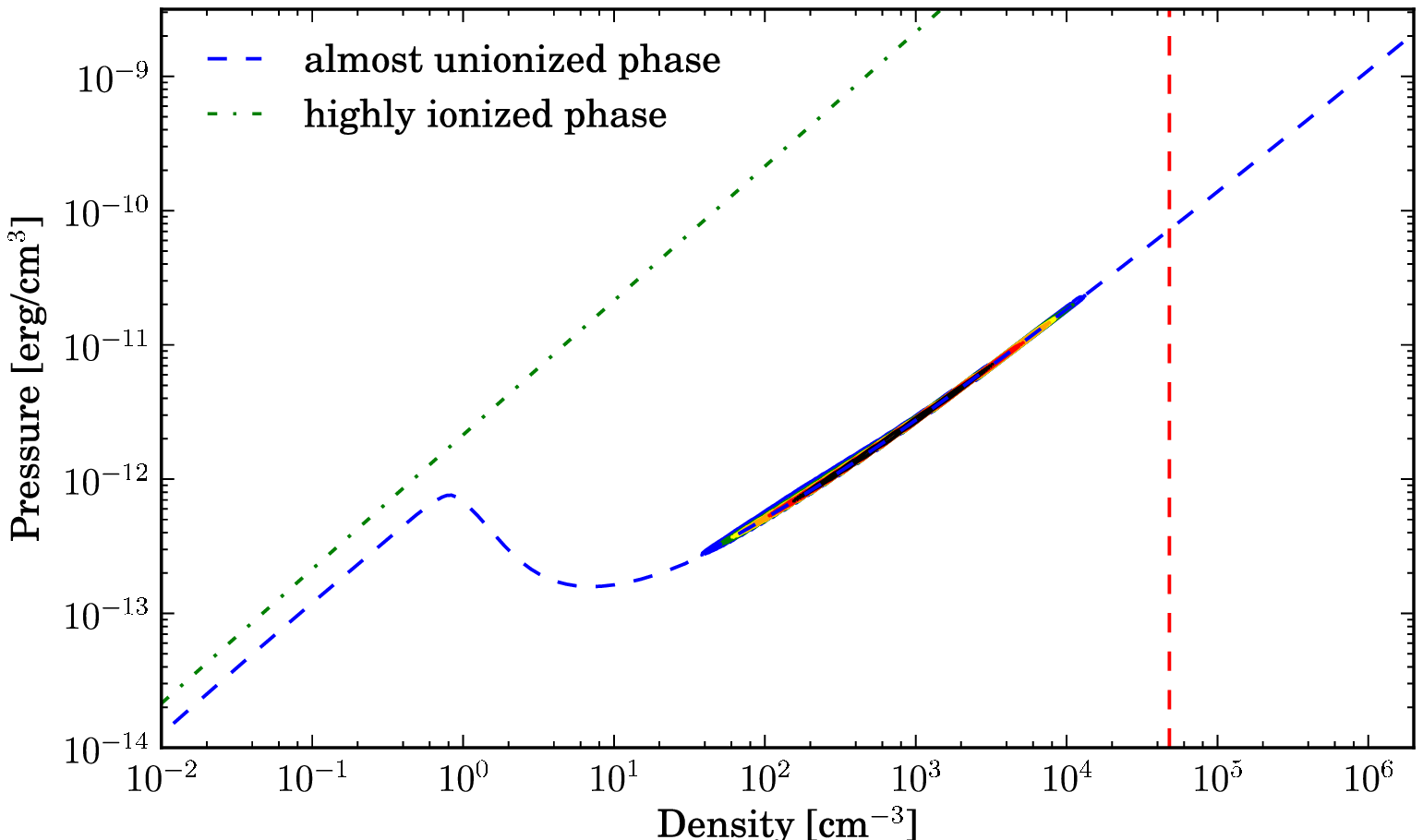}
\includegraphics[width=\linewidth]{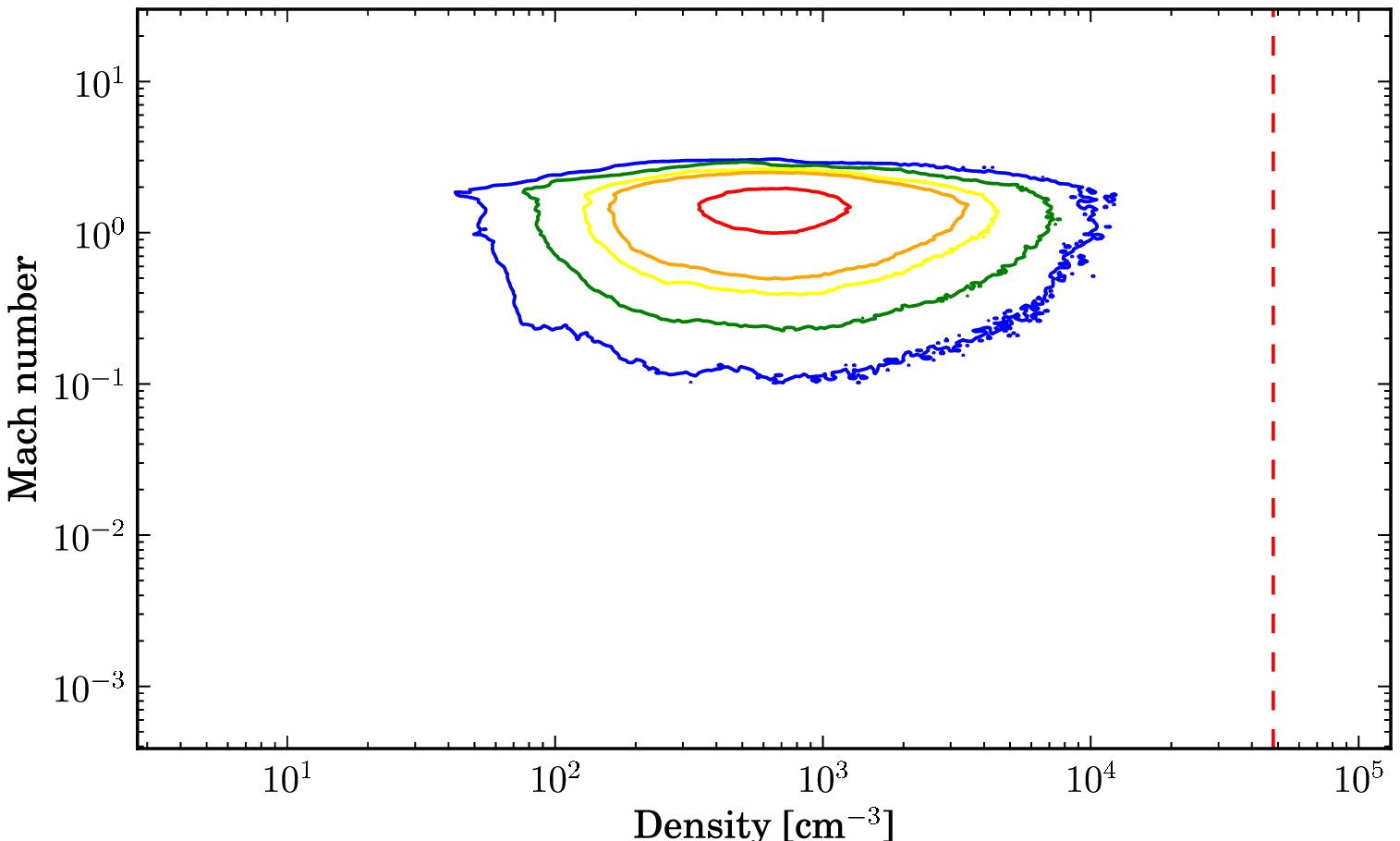}
\caption{\label{histo_tur01_30} 2D plot of the mass fraction at a
  given density and pressure (top) and at a given density and Mach
  number (bottom), before ionization is turned on in the simulation
  with a mean Mach number of one. Blue to black contours are
  increasing mass fraction contours (blue: 1E-6, green: 1E-5, yellow:
  5E-5, orange: 1E-4, red: 5E-4, black: 2E-3). The dashed-red line at
  4.8$\times$10$^4$ /cm$^3$ is the maximum density achieved in a
  plan-parallel 1D simulation for a homogeneous density at 500
  /cm$^3$.} 
\end{figure}

\begin{figure}[!h]
\centering
\includegraphics[width=\linewidth]{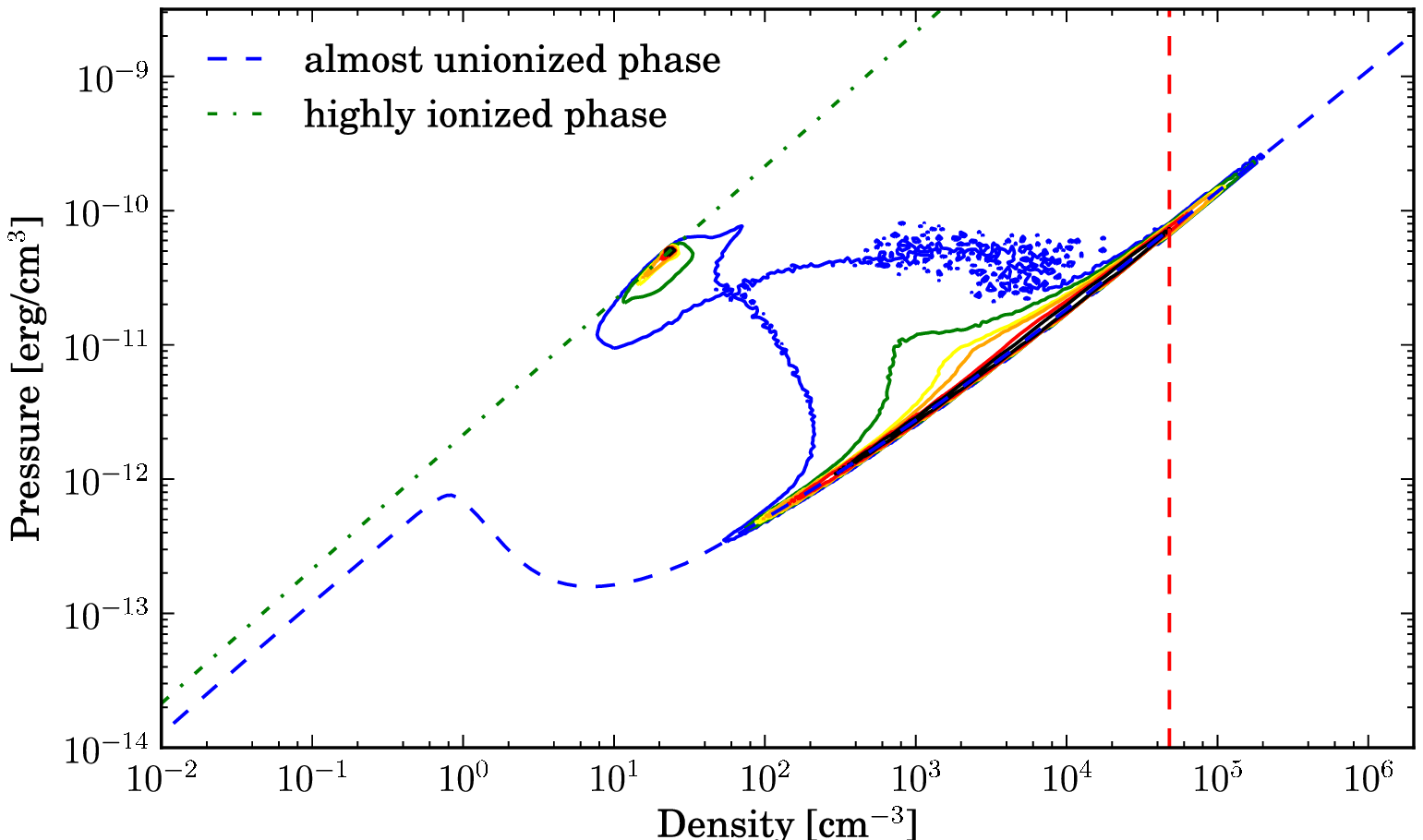}
\includegraphics[width=\linewidth]{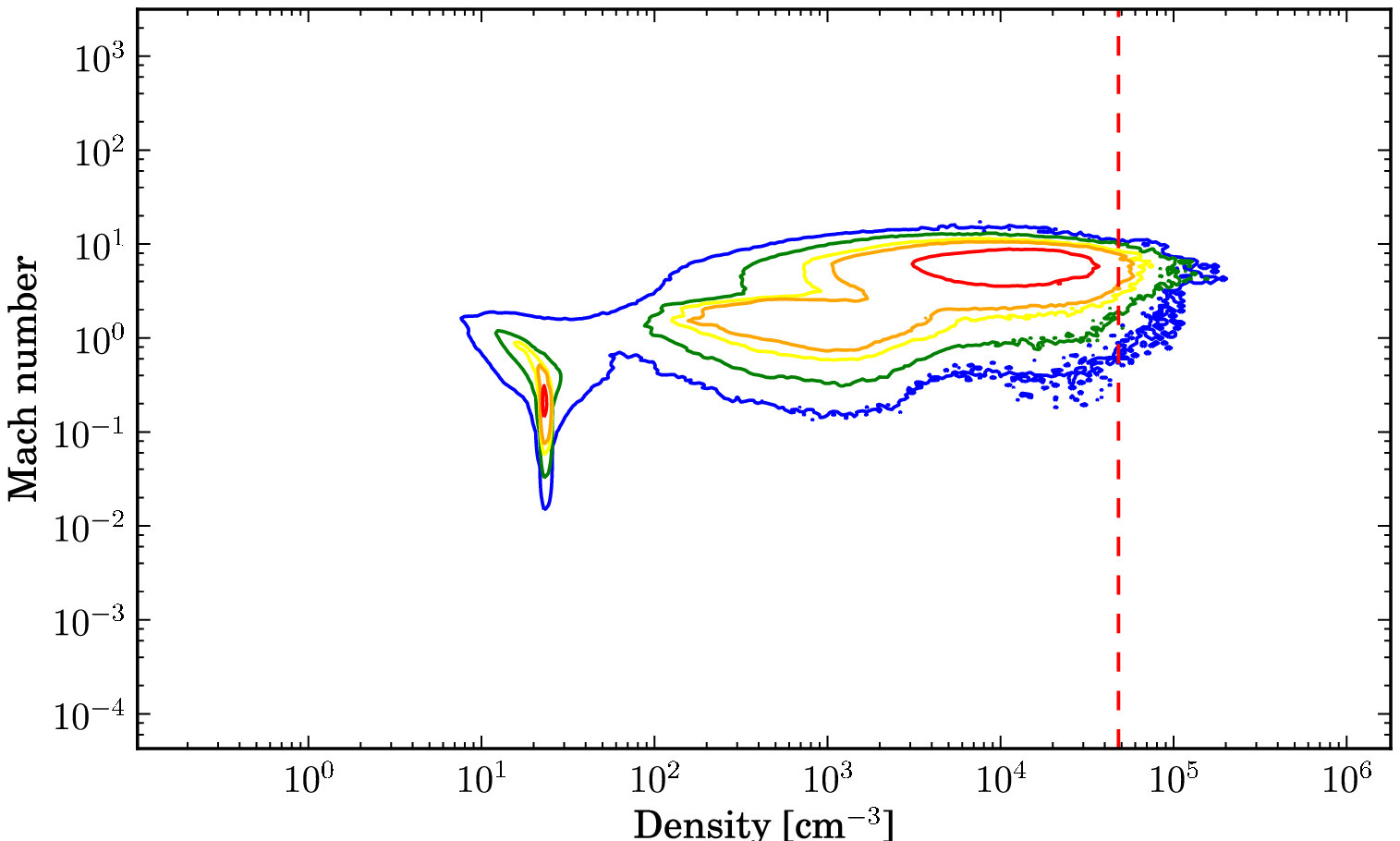}
\caption{\label{histo_tur01_72} 2D plot of the mass fraction at a
  given density and pressure (top) and at a given density and Mach
  number (bottom), 1 My after ionization is turned on. Blue to black
  contours are increasing mass fraction contours (blue: 1E-6, green:
  1E-5, yellow: 5E-5, orange: 1E-4, red: 5E-4, black: 2E-3). The
  dashed-red line is the maximum density achieved in a plan-parallel
  1D simulation for a homogeneous density at 500 /cm$^3$:
  4.8$\times$10$^4$ /cm$^3$.} 
\end{figure}

The first line of Fig. \ref{crho} presents three snapshots of the
column density for the transonic simulation, when ionization is turned
on, 500 ky and 1 My after ionization has started. The hot ionized gas
expands in the cold turbulent medium and triggers the propagation of a
shock ahead of the ionization front. This is the collect part of the
classical collect and collapse scenario \citep{Elmegreen:1977iq} and
is observed in many regions
\citep[see][]{Zavagno:2010jv,Thompson:2011vr,Deharveng:2010cp}. The
500-ky snapshot clearly shows a three-phase medium with the hot
ionized gas at the top, the shocked dense region and the unperturbed
cold cloud. The shocked region is apparently 0.5-1 parsec wide at 500
ky and gets wider in time, up to 1.5 parsec at 1 My. This width is a
lot larger than one would expect from the collect process in a
  homogeneous medium. The typical width of the dense shell is 0.1
parsec in this case. However Fig. \ref{cut_rho} clearly shows that the
real thickness of the shell is indeed of order 0.1 parsec but contrary
to the collect process in a homogeneous medium, the shell is not
flat but highly curved. Therefore the column density plots show a wide
shocked region just by projection effects of this highly turbulent
shocked region.\\ 
\indent The properties of these phases can be studied using the
distribution of the mass fraction in the density-pressure plane or in
the density-Mach number plane. Fig. \ref{histo_tur01_30} and
\ref{histo_tur01_72} show the 2D plots of the mass fraction before and
after ionization is turned on. Before
ionization is turned on,  the distribution of the gas is at thermal
equilibrium and the average Mach number is of order 1 with a
dispersion of 0.48. After 1 My, the three-phase medium is formed and
the three phases are well separated in the Mach number density
plot. The hot ionized gas can be seen on the pressure density plot, on
the isothermal equilibrium curve for the ionized gas, and on the Mach
number density plot in the sub-sonic area. The unperturbed medium
remains at the same location in the graph with respect to
Fig. \ref{histo_tur01_30} while the shocked material is pushed at high
Mach number and high densities. The distribution of the mass fraction
is quite similar to the one found in paper I, the only difference is
the initial distribution of the gas, which was previously homogeneous
at 500 cm$^{-3}$ with a density or interface modulation. In both
cases, the shocked region contains points with densities
  higher than the one corresponding to the collect and collapse
  scenario (red dashed line in Fig. \ref{histo_tur01_30} and
\ref{histo_tur01_72}), i.e. the density after the collect phase in a
homogeneous medium. These points represent 3 \% of the shell
  in mass fraction and help to achieve densities high enough to
  trigger the gravitational collapse.

\subsection{Supersonic turbulence and comparison}
\begin{figure*}[!ht]
\centering
\includegraphics[trim=5cm 0.5cm 3cm
  0.5cm,clip,width=\linewidth]{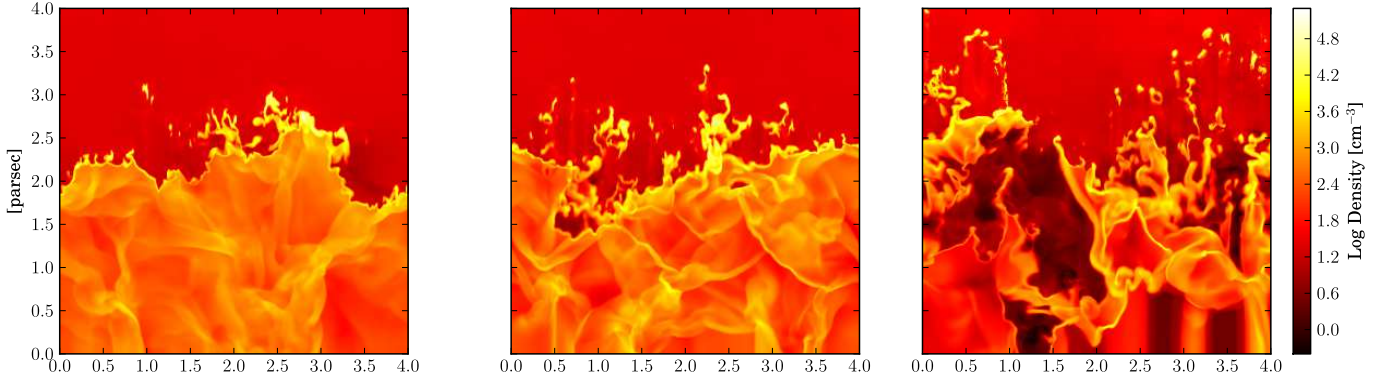}
\caption{\label{cut_rho} Density cuts through the three different
  simulations at t = 500 ky. Left: Mach 1, middle Mach 2 and right
  Mach 4.}
\end{figure*}
\begin{figure}[!ht]
\centering
\includegraphics[width=\linewidth]{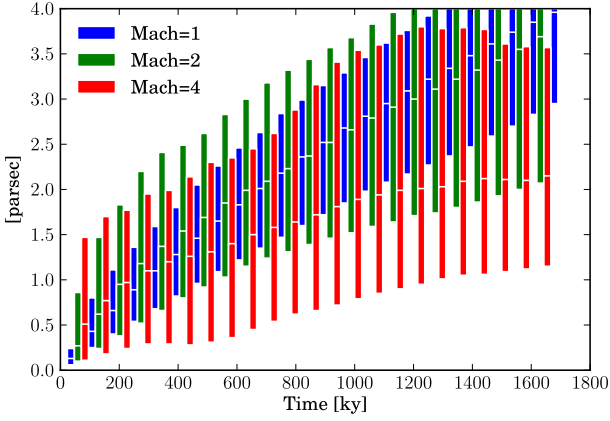}
\caption{\label{ifront} Positions of the 0.95-, 0.5- and
  0.05-transition of the xy-average ionization fraction. Between 0 and
  the lower bound the ionization fraction of the gas is between 1 and
  0.95, between the lower bound and the middle bound, between 0.95 and
  0.5 and between the middle bound and the upper bound the ionization
  fraction is between 0.5 and 0.05. } 
\end{figure}

The simulations with a mean Mach number of two and four present the
same three phase structures: the ionized region, the shocked region,
and the unperturbed region. However, in these cases the initial gas
has more turbulent ram pressure to resist the ionized-gas
expansion. The pressure of the ionized gas and the ram pressure caused
by turbulence can be estimated from 
\begin{eqnarray}\label{pressures}
p_{II}   &=& 2\bar{n}_{II}k_b T_{II}\cr
p_{turb} &=& \overline{\rho_0 v_{0}^2}
\end{eqnarray}
where $n_{II}$ and $T_{II} \approx 7700$ K are the density and the
temperature in the ionized gas and where $\rho_0$ and $v_{0}$ the
density and the mean velocity in the initial turbulent medium. The
density of the ionized gas is nearly constant in time in the three
simulations, and does not depend on the initial turbulence. $n_{II}$
is around 10/cm$^3$ and the corresponding pressure at t = 500 ky is given in Table
\ref{pressures_param} with the initial turbulent ram
pressure. The ionized-gas pressure dominates the turbulent ram
pressure for a mean Mach number of one and two. The consequent
progression of the ionized-gas expansion is clearly visible in the
column density plots in Fig. \ref{crho}. However, the plots also show
that this progression is nearly stopped for a mean Mach number of
four. This is caused by the fact that the turbulent ram pressure dominates the ionized-gas pressure (see Table
\ref{pressures_param}). Another interesting difference between the
Mach-four case and the others is the presence of many globules. It
seems that the presence of these globules in our simulations can be
linked with the fact that turbulent motions dominate the ionized-gas
expansion, this will be investigated in more detail in
Sect. \ref{structures}.\\ 
\begin{table}[!ht]
\caption{\label{pressures_param} Initial turbulent ram pressure and
  ionized-gas pressure at t = 500 ky} 
\centering
\begin{tabular}{l|l|l|l}
Mean Mach number & 1 & 2 & 4 \\
\hline
$p_{turb}/kb$ at t = 0 ky [K/cm$^3$] & 3.2$\times$10$^4$ &
1.2$\times$10$^5$ & 4.5$\times$10$^5$ \\ 
$p_{II}/kb$ at t = 500 ky [K/cm$^3$] & 1.7$\times$10$^5$ &
1.9$\times$10$^5$ & 1.8$\times$10$^5$ \\ 
\end{tabular}
\end{table}
\indent Locally the ionization fraction changes from one in the
ionized region to zero in the unionized region in a very small zone
(less than 0.01 pc) when turbulence can be neglected. Therefore the
position at which the fraction is less than 0.5 gives the position of
the ionization front. When turbulence is high, the position of the
front is difficult to find since there is mixing between ionized gas
and cold gas (see Fig. \ref{cut_rho}). However global parameters like
the mean position and apparent width of the ionizing front can still
be studied thanks to the whole ionization fraction field. We used an
average of the vertical profile of the ionization fraction. It
transits between one in the completely ionized region and zero in the
cold region, in between it slowly decreases because of the apparent
width caused by the projection. We define the mean position of the
ionizing front $Z_X$ where the average profile is at 0.5 and the width
$\Delta Z_{X}$ as the distance between the 0.95 and 0.05
positions. This corresponds to the following equations 
\begin{eqnarray}
Z_{X}& = &\int^{L_{box}}_{0}H(\langle X(x,y,z)\rangle_{x,y}- 0.5)dz, \cr
\Delta Z_{X}& = &\int^{L_{box}}_{0}H(\langle X\rangle_{x,y} -
0.05)-H(\langle X\rangle_{x,y} -0.95)dz, \cr 
H(x - y)& = &1\textrm{, if } x\geq y \textrm{ or } 0\textrm{, if } x<y.
\end{eqnarray}
where $X(x,y,z)$ is the ionization fraction field and $H$ the
Heavy-side function. $Z_X$ and $\Delta Z_{X}$ are shown in
Fig. \ref{ifront}. The front progression is slowed down and its width
increases as the Mach number increases. This is caused by the bigger
density contrast with increasing turbulence. In the three cases, the
width of the shell is mainly caused by a projection effect of the
column density. The real shell is thin (see Fig. \ref{cut_rho}), but
is very spatially disperse by the inner structures of the
cloud. Therefore, it will look wider in column density. \\


\subsection{Ionization and thermal equilibrium}
\begin{figure}[!ht]
\centering
\includegraphics[width=\linewidth]{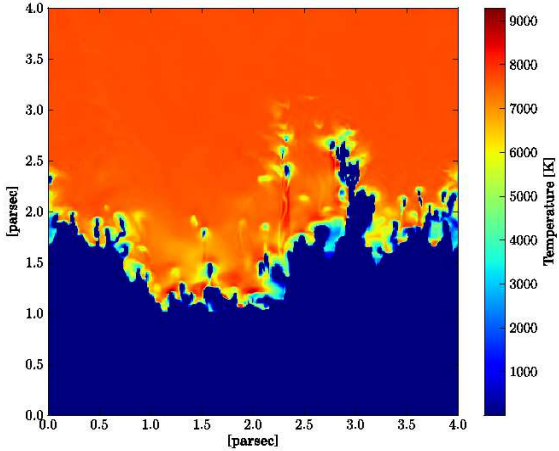}
\caption{\label{cut_T} Temperature cut at t $\approx$ 710 ky after
  ionization is turned on. The 3D temperature effects studied in the
  clump simulation (sect. \ref{num_met} are clearly visible through
  the snapshot.} 
\end{figure}

The scheme we used (Sect. \ref{num_met}) is able to treat the
ionization physics out of equilibrium. Therefore an interesting
quantity to follow is the fraction of gas which is out of
equilibrium. At equilibrium, the ionization rate compensates
  the recombination rate. From Eq. \ref{io_eq}, the ionization
  fraction at equilibrium is $X=(\sqrt{1+4/y}-1)y/2$ with
  $y=\sigma_\gamma F_\gamma/n_H \beta$. In the molecular cloud
  $F_\gamma$ is 0 therefore X is 0 at equilibrium. In the HII region,
  $y \gg 1$ so the ionization fraction tends to 1. At the ionization
  front, the ionization fraction can be at equilibrium and in between
  0 and 1, however this region is very small and can be neglected in
  the simulation (less than 1 \% of the box). The thickness of the
  ionization front is given by $1/n_H\sigma_\gamma$ which is of the
  order of 10$^{-4}$ pc for $n_H$=500/cm$^3$. Therefore we define the
fraction of gas out of equilibrium as the percentage of cells that
have a ionization fraction between 0.05 and 0.95, the values at
different times for the three simulations are given in Table
\ref{out_eq}. The more turbulent the medium, the more out of
equilibrium. It is interesting to note that this is similar to
  what was observed with the thermal instability in the interstellar
  medium
  \citep[see][]{SanchezSalcedo:2002fp,Gazol:2005go,Audit:2005df}. Almost
one fifth of the simulation box at t = 1 My is out of equilibrium for
a mean Mach number of four. This means that the hypothesis of an
ionization equilibrium is quite bad to study the impact of ionization
on a turbulent cloud. This is mainly caused by the ionized gas getting
into a shadowed region. In these regions the recombination time is
long, between 1 and 100 ky for densities around 1.-10. /cm$^3$ and
temperature around 8000 K as we have already explained in
sect. \ref{tests_sect}.\\ 
\begin{table}[!ht]
\caption{\label{out_eq} Percentage of cells out of ionization
  equilibrium.} 
\centering
\begin{tabular}{l|l|l|l}
Time & Mach 1 & Mach 2 & Mach 4 \\
\hline
 250 ky & 1 \% & 7 \% & 8 \% \\
 500 ky & 3 \% & 9 \% & 12 \% \\
 750 ky & 5 \% & 10 \% & 13 \% \\
   1 My & 6 \% & 12 \% & 17 \% 
\end{tabular}
\end{table}
\indent The temperature is also an important quantity to monitor in
our scheme. We solve the thermal balance between ionization and
recombination and this allows the gas to be at states out of
equilibrium as shown in sect. \ref{num_met}. This behavior can be
identified in Fig. \ref{cut_T}, the gas ahead of the dense parts
facing the ionization flux is ionized but at a temperature between
2000 and 6000 K. This is caused by the 3D expansion of the dense
ionized gas at the top of the structures facing the ionization
radiation. Besides, the regions that are in the shadow of the dense
structures tend to be overheated at almost 10 000 Kelvin whereas the
equilibrium temperature is around 7750 Kelvin. This is caused by the
fact that the recombination process in the shadow is done at constant
pressure, the pressure being imposed by the hot ionized and exposed
gas surrounding the shadowed gas. Therefore, as we have demonstrated
in sect. \ref{num_met} for the simple situation of a clump shadowing
parts of the hot ionized gas, the temperature behind the clump will
increase. In the turbulent simulation, the cooling processes for the
unionized gas is also playing a role (see paper I for details), they
will cool the recombined gas so that the temperature does not reach 14
000 Kelvin but equilibrates around 10 000 Kelvin. This result is
counter-intuitive since one would expect that the gas stays at 7750
Kelvin during the recombination step, the kinetic energy of the
electrons being radiatively lost for the system. However, the 3D
versus 1D study done in sect. \ref{num_met} clearly show the
importance of the lateral motions from the exposed gas to the shadowed
gas to impose a recombination at constant pressure behind the
structures. \\    


\section{Structures and observational signatures}\label{structures}

\subsection{Pillars}

\begin{figure}[h]
\centering
\includegraphics[width=\linewidth]{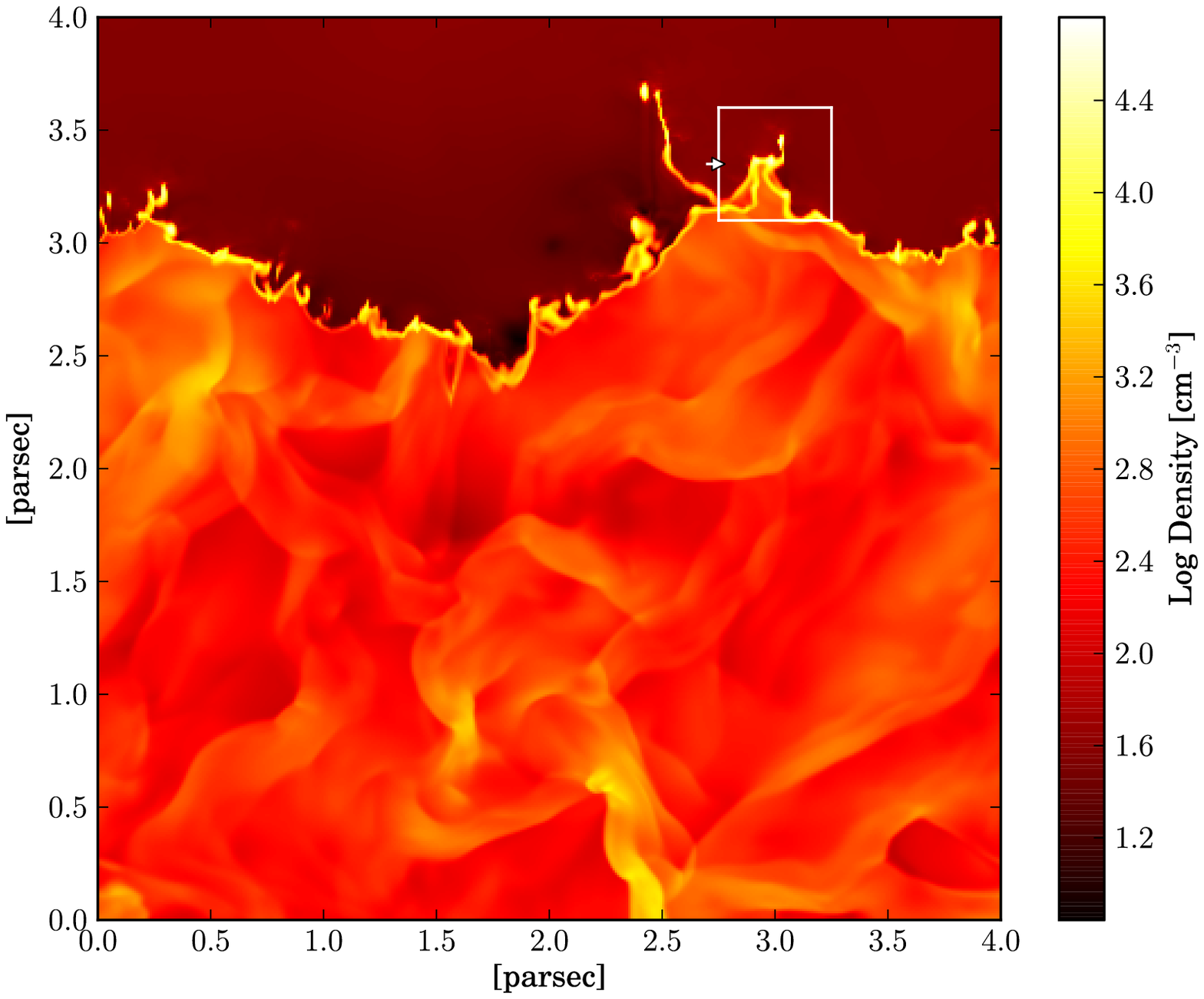}
\includegraphics[width=\linewidth]{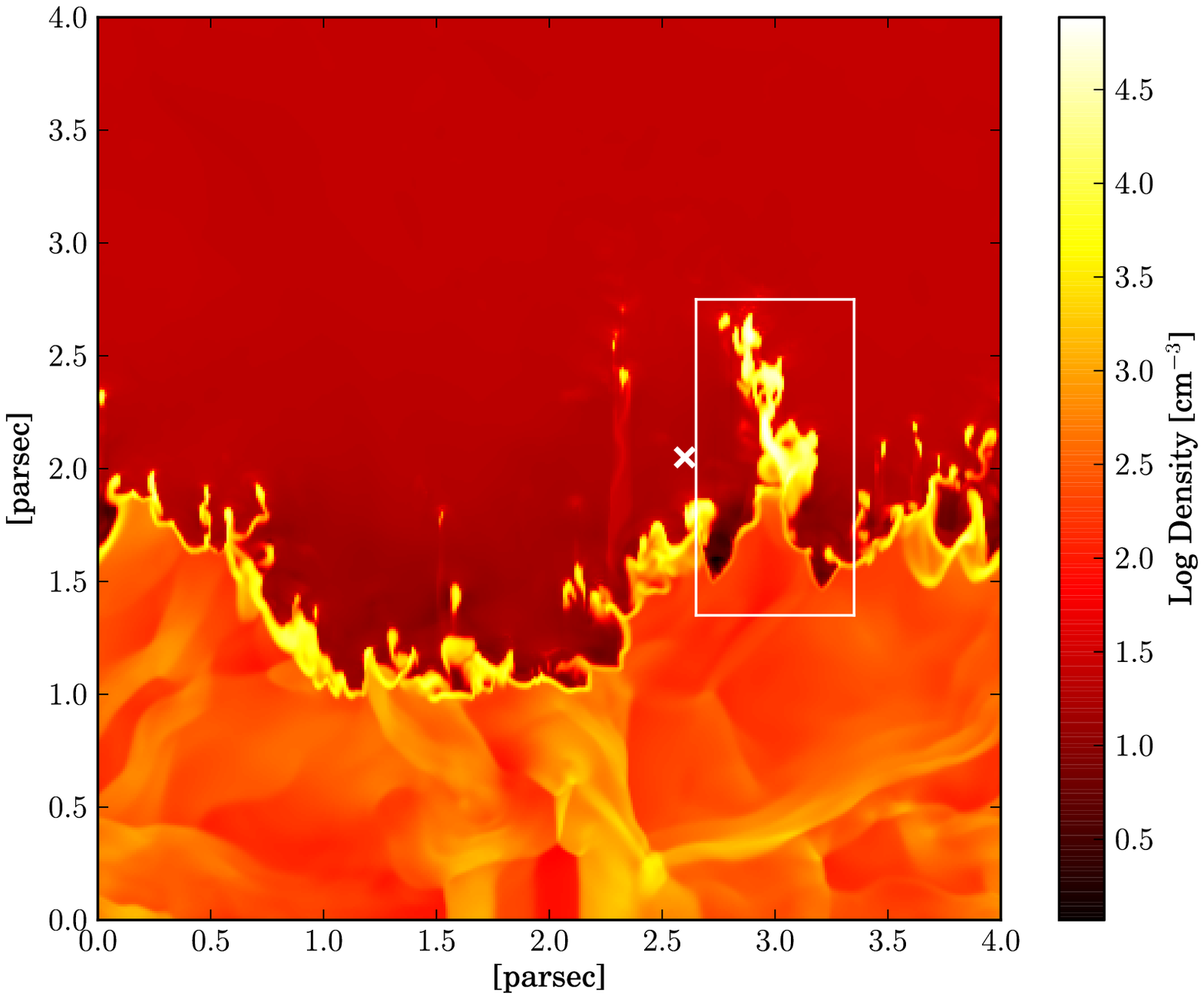}
\caption{\label{cut_rho_tur01} Density cut at t $\approx$ 240 ky and t
  $\approx$ 710 ky after ionization is turned on. The white box
  indicates the areas of the column density and line-of-sight velocity
  spectra made in Fig. \ref{los_30} and \ref{los_60}. The arrow
  indicates the direction of the line of sight; for the cross, the
  line of sight is perpendicular to the density cut plane. } 
\end{figure}

\begin{figure}[h]
\centering
\includegraphics[width=\linewidth]{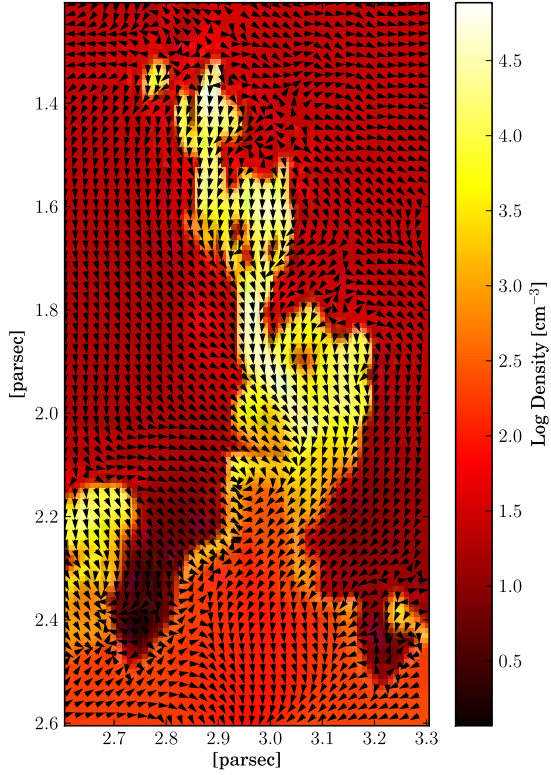}
\caption{\label{velocity_field} Cut of density field and velocity
  field orientation along the pillar at t $\approx$ 710 ky.} 
\end{figure}

\begin{figure}[h]
\centering
\includegraphics[trim=1.8cm 3.3cm 1.5cm
  3.3cm,clip,width=0.45\linewidth]{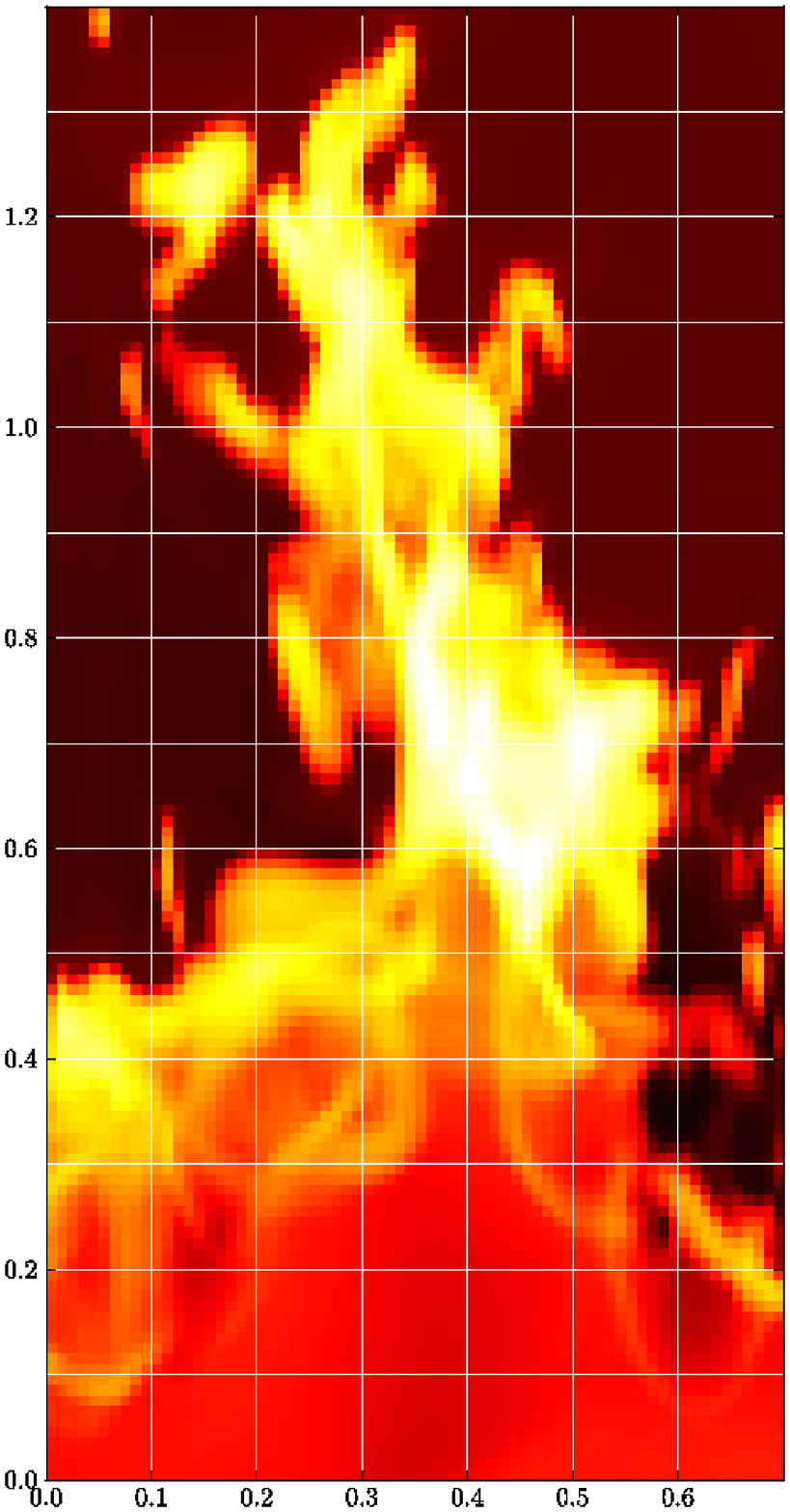} 
\includegraphics[trim=1.8cm 3cm 1.5cm
  3cm,clip,width=0.45\linewidth]{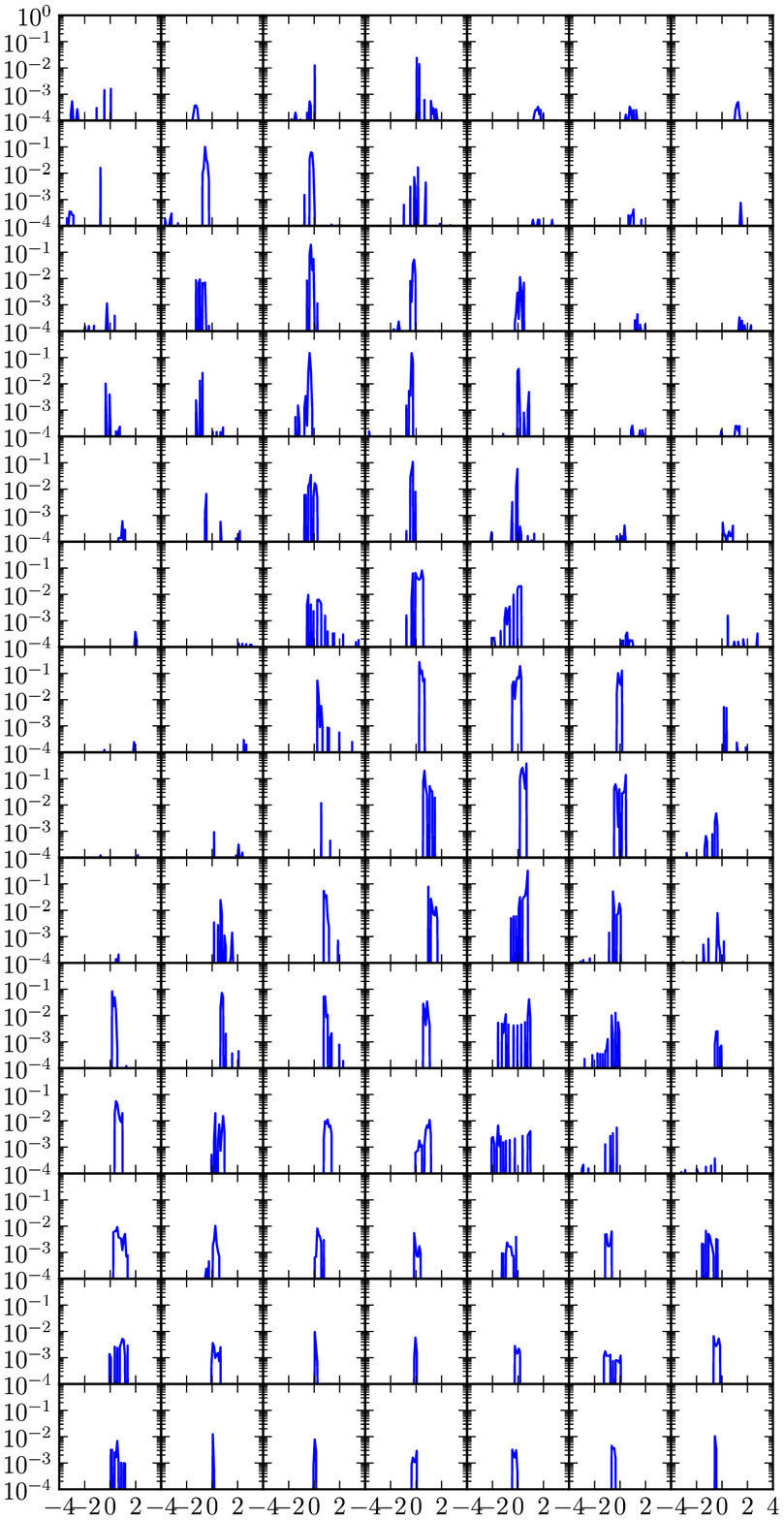} 
\caption{\label{los_60} Left: column density on a 0.7x1.4x0.5 pc$^3$
  box around the pillar structure at t $\approx$ 710 ky. Right:
  mass-weighted histogram of the line of sight velocity in the same
  box (similar to optically-thin observational line spectra). Each
  spectrum is made on a square of 0.1x0.1 pc$^2$ drawn on the column
  density map. The spectra are drawn between -4 and 4 km/s in 80 bins
  (horizontal axis) and the mass between 10$^{−4}$ and 1 solar mass
  (vertical axis in log scale). The lateral shocks have collided to
  form the pillar, the spectra are now peaked around 0 km/s.} 
\end{figure}

It is not clear how to distinguish structures on the column density
maps in Fig. \ref{crho} because of the confusion on the line of
sight. Therefore we will investigate some density cuts to identify
pillars and globules. Figure \ref{cut_rho_tur01} presents the
formation of a pillar in the Mach 1 simulation between 200 and 700
ky. The pillar formed is one parsec long and presents a complex
structure, with successive dense parts along its length. Figure
\ref{velocity_field} is a map along the pillar with the direction of
the velocity field. The structure of the field is similar to the one
already obtained in paper I for non-turbulent set-ups. The pillar has
two dense heads at a vertical position of 1.4 and 1.6 parsec, a dense
base between 1.8 and 2 parsec and two holes on the side at 2.4
parsec. The velocity field shows that the pillar is forming at its
base by the convergence of the dense parts of the shell while the
holes are formed by the divergence of the dense parts created on a
concave zone. This is the same configuration that was previously seen
in paper I, without any turbulence. This suggests that the mechanism
at the origin of the formation of pillars are not to much dependent on
the turbulence. Here the turbulence gives only the initial conditions
in terms of density contrasts and initial structures on which pillars
will form. The holes on the side of the pillars in the non-turbulent
set-ups of paper I were in fact a complete annulus around the pillars
caused by the symmetry of our initial conditions. The major difference
with a turbulent scenario is that there is no more symmetry in the
initial conditions. This leads to holes that are local and do not
extend in a full annulus. \\ 
\begin{figure}[h]
\centering
\includegraphics[trim=1.8cm 2.6cm 1.5cm
  2.6cm,clip,width=0.45\linewidth]{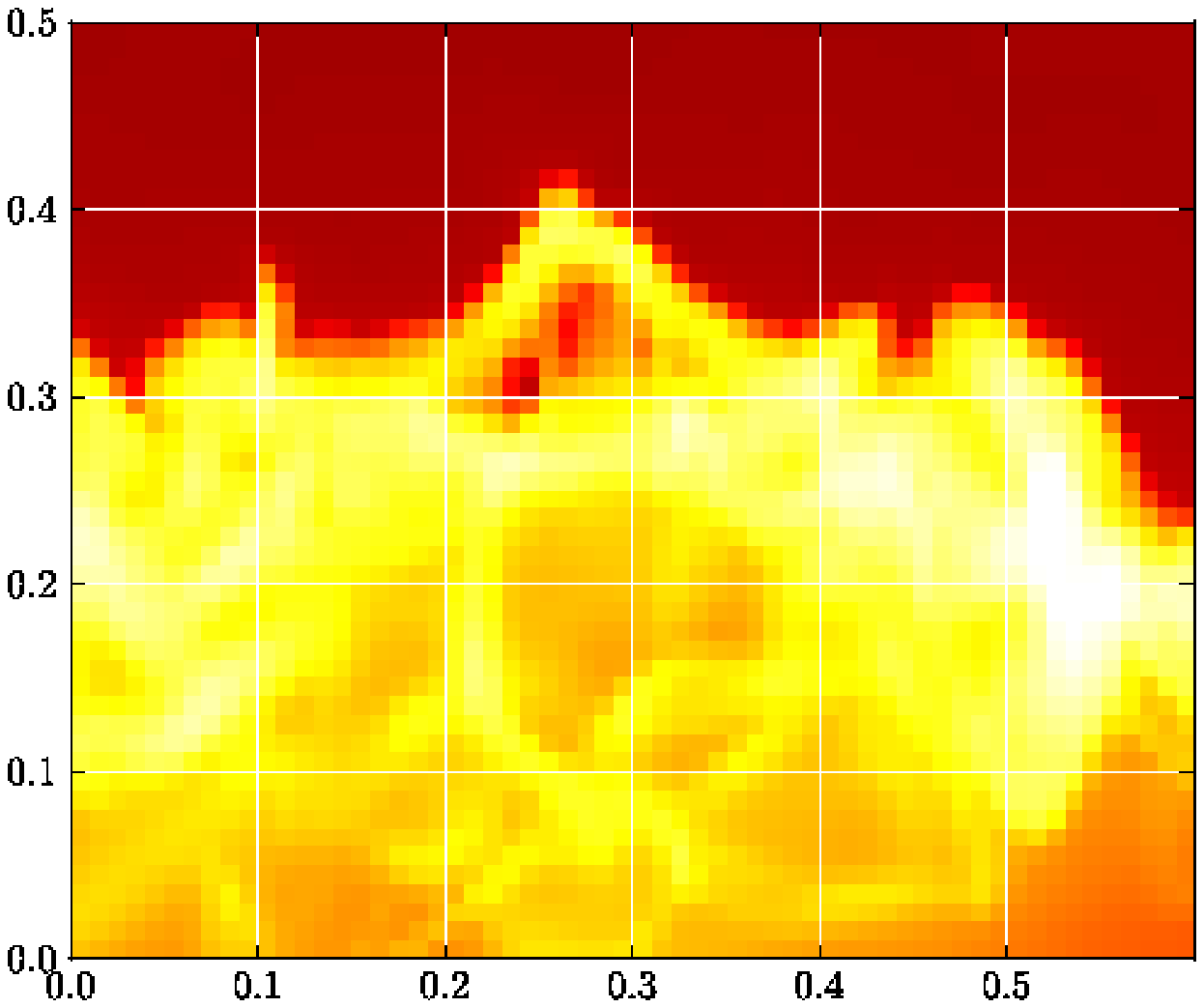} 
\includegraphics[trim=1.8cm 1.5cm 1.5cm
  1.5cm,clip,width=0.45\linewidth]{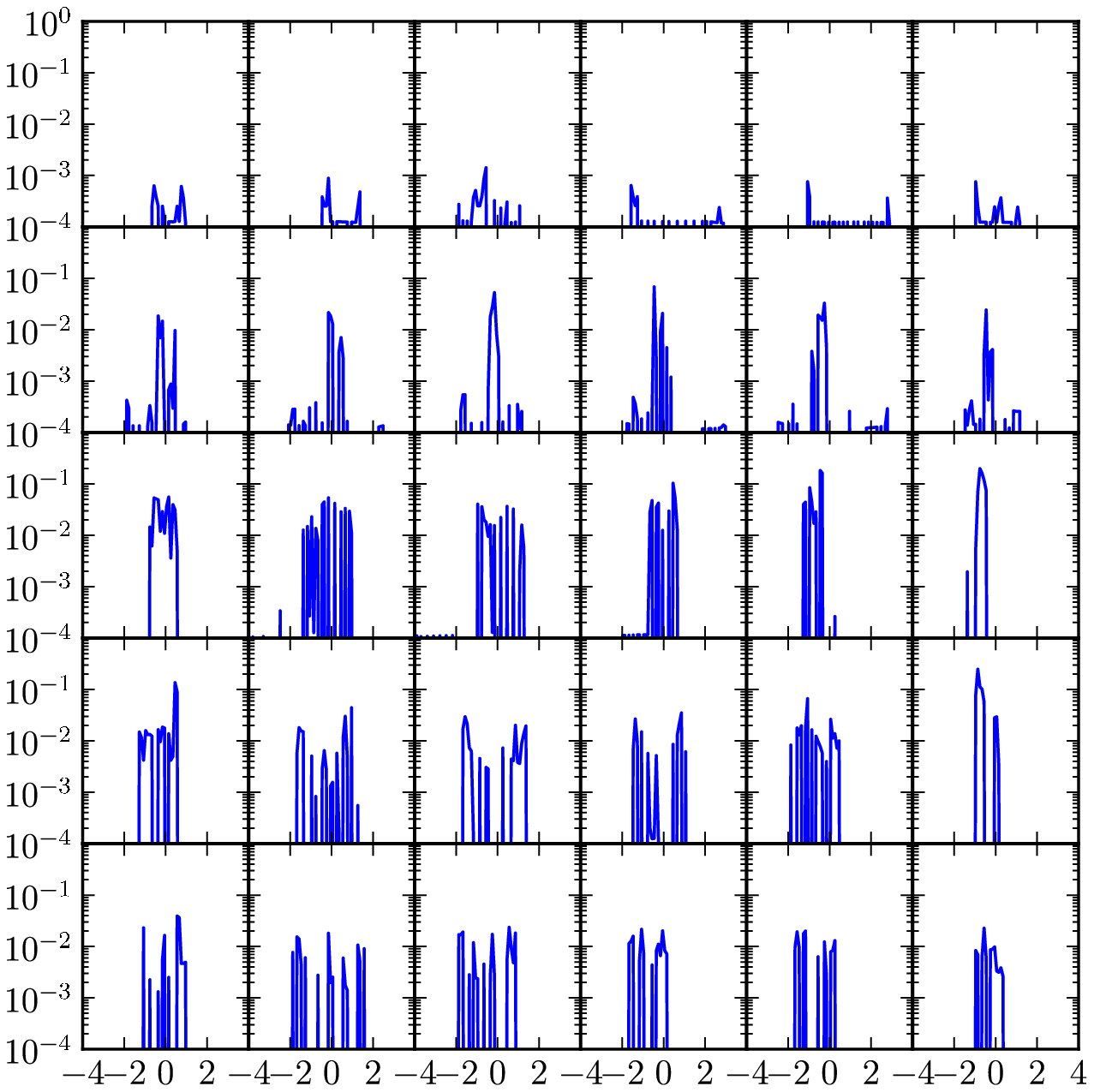} 
\caption{\label{los_30} Left: column density on a 0.6x0.5x0.5 pc$^3$
  box around the area in which the pillar will form at t$\approx$ 240
  ky. Right: mass-weighted histogram of the line of sight velocity in
  the same box (similar to optically-thin observational line
  spectra). Each spectrum is made on a square of 0.1x0.1 pc$^2$ drawn
  on the column density map. The spectra are drawn between -4 and 4
  km/s in 80 bins (horizontal axis) and the mass between 10$^{−4}$ and
  1 solar mass (vertical axis in log scale). The lateral shocks can
  be identified in the wide spectra leading to a very broad
    averaged line width.} 
\end{figure}
\indent The curvature of the shell can also be seen on spectra of the
line-of-sight velocity. Figures \ref{los_30} and \ref{los_60} show on
the left a column density maps of the pillar and on the right the mass
histograms as a function of the line-of-sight velocity in squares of
0.05 parsec. Figure \ref{los_30} is a snapshot at t $\approx$ 240 ky
when the pillar is not yet formed and Fig. \ref{los_60} at t $\approx$
710 ky with the whole pillar. The line-of-sight velocity spectra are
wide (between -2 and +2 km/s) before the pillar is
formed. This was already identified in paper I as the signature of the
dense shell which is curved on itself with the two components,
blue-shifted and red-shifted, that are going to collide to form the
pillar. When the collision has occurred, the wide spectra is no longer
visible and the line-of-sight velocity spectra are peaked around a
null velocity (see Fig. \ref{los_60}). This demonstrates that the
scenario identified in paper I with the study of non-turbulent set-ups
is also taking place in our turbulent simulation. \\ 
\indent Based on this analysis, we can infer that molecular tracers
that are optically thin should reflect the same spectral structure as
the line-of-sight velocity spectra of the simulations. Therefore, wide
spectra could be the indication of nascent pillars whereas evolved
pillars should present a spectra peaked around the velocity of the
expanding bubble projected on the line of sight. This velocity is zero
in our case since the line of sight is perpendicular to the direction
of the expansion. \\ 

\subsection{Dense clumps at the edge of the ionized gas}

\begin{figure}[h]
\centering
\includegraphics[width=\linewidth]{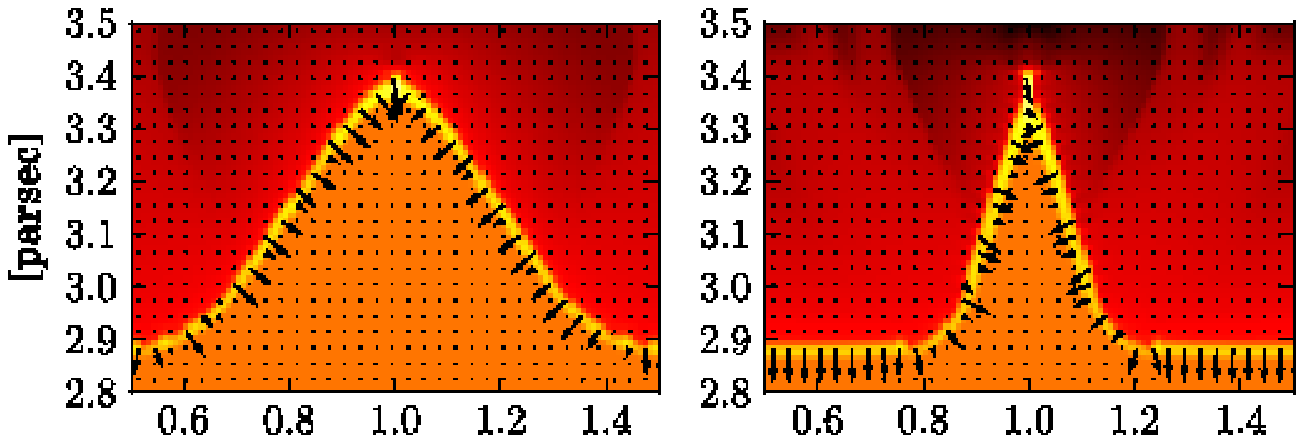} 
\includegraphics[width=\linewidth]{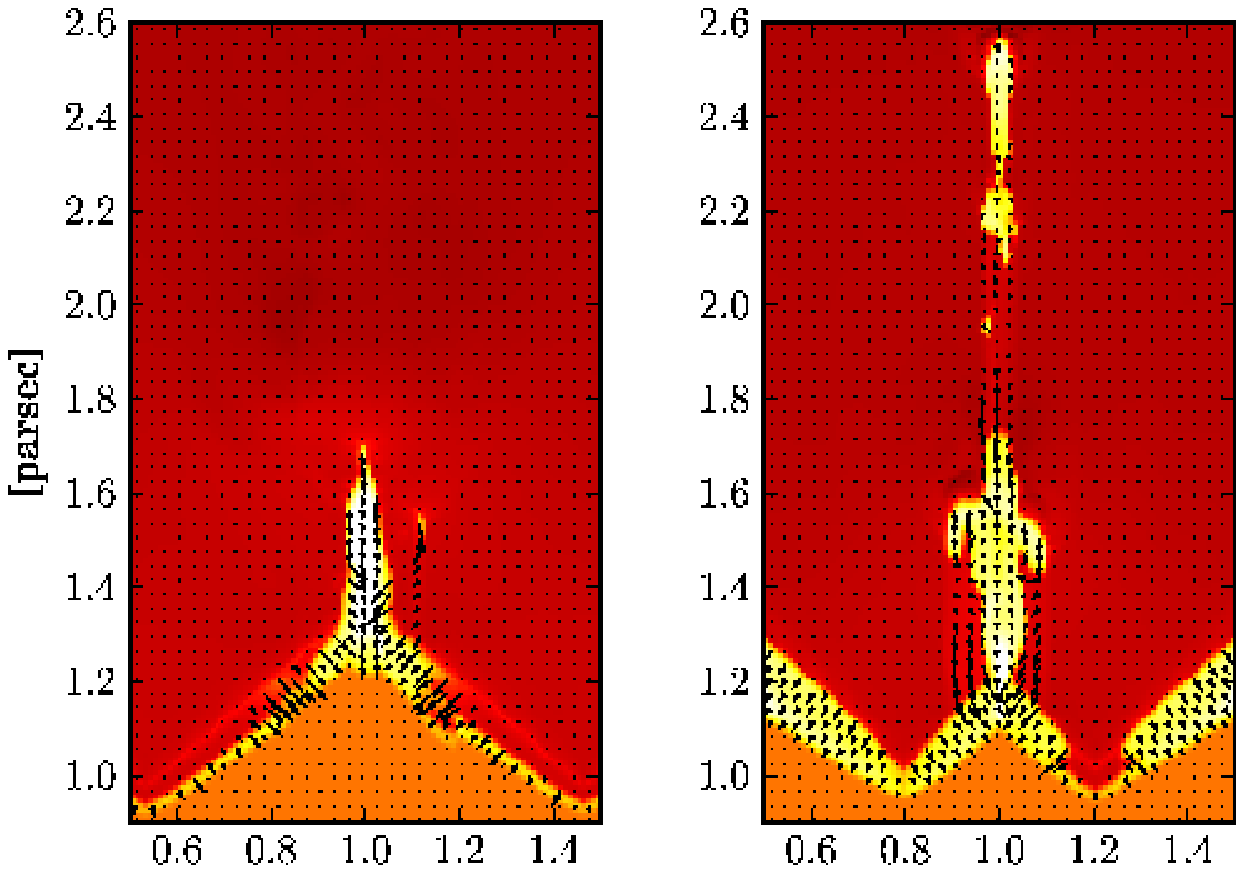} 
\caption{\label{fig_collapse} Illustration of the effect of the shell
  curvature on the formation of clumps and pillars. Density cut and velocity field
  (for the unionized gas) are taken from simulations from paper I. Initial curvature radius:
  left: 5$\times$10$^{-2}$ pc and right: 1$\times$10$^{-2}$ pc. Top:
  20 ky after the beginning 
  of the simulation, the shell is formed on the curved
  surface. Bottom: snapshots after 550 ky. A 1.5-pc long pillar is
  formed at high curvature whereas a clump of the size of the initial
  structure is formed at low curvature.}
\end{figure}

The difference between the formation mechanism of pillars and clumps
inside the shell is also clearly visible in
Fig. \ref{cut_rho_tur01}. On the first snapshot the position on which
the pillar is forming can be identified as the position at which the
curvature of the shell is higher. On the other parts of the shell, the
curvature is not sufficient to form pillars but will rather trigger
the formation of clumps inside the shell. This phenomenon was already
seen in Paper I, the higher the curvature, the longer the
pillar. If the shell is curved enough to collapse on itself it
  will form a pillar. At high curvature, the tip of  the curved shell
  will be stopped by the collapse of the matter ahead of it, which leads
  to the formation of a pillar (Right part of
  Fig. \ref{fig_collapse}, at early times the velocity field of the
  shocked gas shows that the shell collapses on the centre of the
  initial structure). The
 shell around the ``hill'' will collapse quickly on the hill, with velocities
 that are nearly perpendicular to the direction of the expansion. That
 is why the motion of the gas at the top of the hill is small. When
 the curvature is low, the shell collapse on the hill but with a velocity
 that has a component in the direction of the expansion. Therefore
 after the collapse, the shocked gas keeps a velocity in the direction of
 the expansion and propagates with the rest of the shell.
  The matter is accelerated with the shell and the density
  increases because the gas surrounding it converges smoothly at the
  centre (Left part of Fig. \ref{fig_collapse}). This situation was
studied and compared in the context of 
RCW36 in Vela C molecular cloud (see Minier et al. 2012 in prep). The
new observations from the Herschel space telescope show the presence
of dense clumps around the HII region RCW36. Dedicated numerical
simulations show that it is rather unlikely that these clumps were
pre-existing because they would have triggered the formation of
pillars. These dense clumps are located at the edge of the HII region
and therefore they are formed by the lateral flows in the shell caused
by curvature perturbations that are not high enough to form
pillars. It is exactly the same process that is at work in the present
turbulent simulations. From paper I, it can be estimated that
  the transition from clump to pillar formation is at a curvature
  radius between 2.5$\times$10$^{-2}$ and 5$\times$10$^{-2}$ pc in the
  present situation.\\
\indent \cite{Thompson:2011vr} showed that the dense clumps around HII
regions are preferentially located at the interface between the
ionized gas and the cold gas. We propose here a formation mechanism of
these dense clumps in the shell thanks to curvature
perturbations. However if the turbulent ram pressure dominates the
ionized gas pressure, such dense clumps will not be present exactly at
the interface. the cold gas as sufficient kinetic energy to disrupt
the interface and a mixing region between the dense shocked gas and
the ionized gas is present (see Fig. \ref{cut_rho}). Therefore the
clear correlation in the observations between the position of the
dense clumps and the interface suggests that in most of these regions
the ionized-gas pressure dominates the turbulent ram pressure of the
surrounding gas.  

\subsection{Globules}

\begin{figure}[h]
\centering
\includegraphics[trim=2cm 1.6cm 1.6cm
  1.6cm,clip,width=0.45\linewidth]{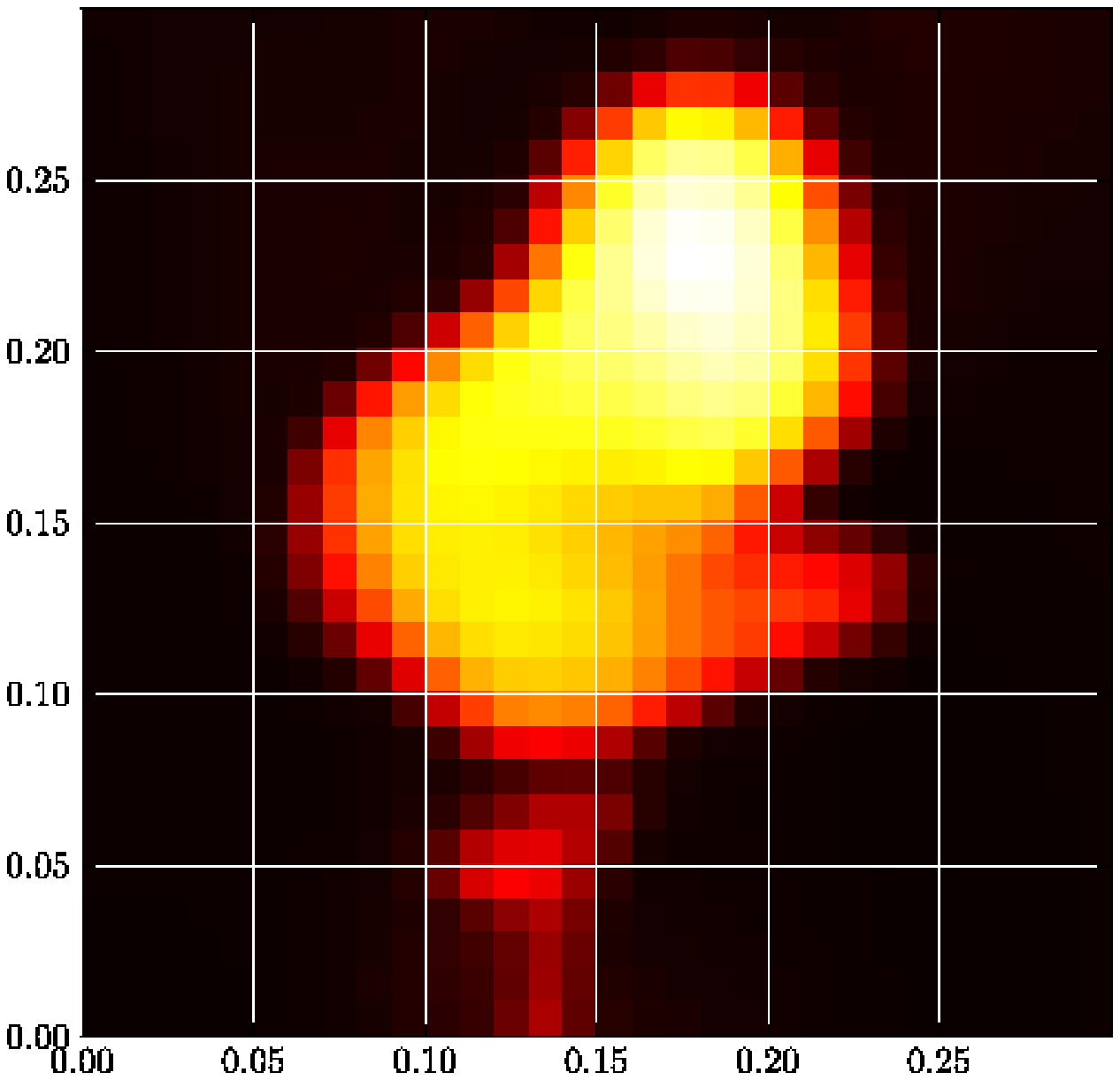} 
\includegraphics[trim=1.9cm 1.5cm 1.5cm
  1.5cm,clip,width=0.45\linewidth]{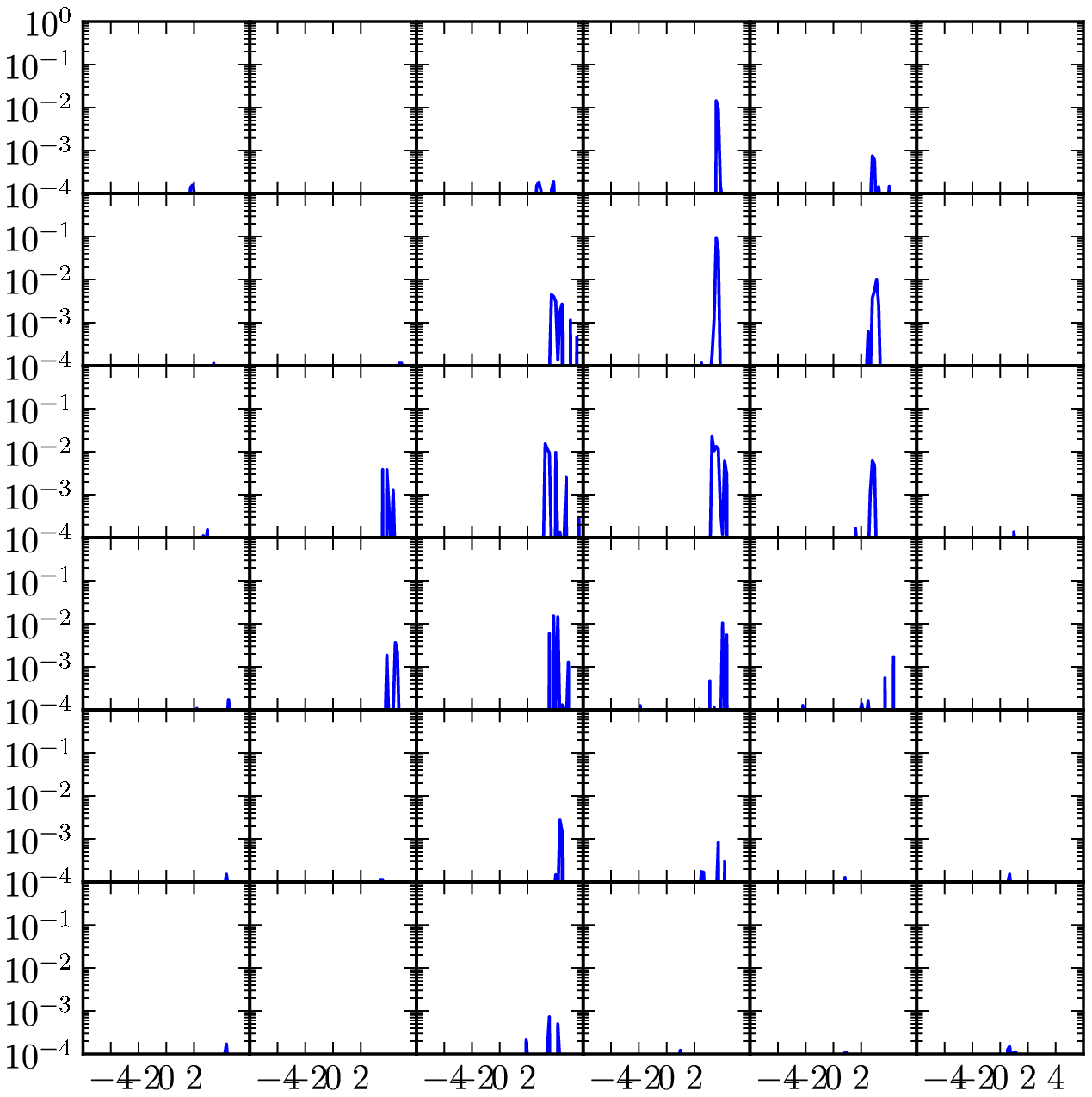} 
\caption{\label{los_globule} Left: column density on a 0.3x0.3x0.25
  pc$^3$ box around a globule in the Mach 4 simulation. Right:
  mass-weighted histogram of the line of sight velocity in the same
  box (similar to optically-thin observational line spectra). Each
  spectrum is made on a square of 0.05x0.05 pc$^2$ drawn on the column
  density map. The spectra are drawn between -6 and 6 km/s in 80 bins
  (horizontal axis) and the mass between 10$^{−4}$ and 1 solar mass
  (vertical axis in log scale). All the spectra are red-shifted at a
  velocity of +4 km/s.} 
\end{figure}

\begin{figure}[h]
\centering
\includegraphics[width=\linewidth]{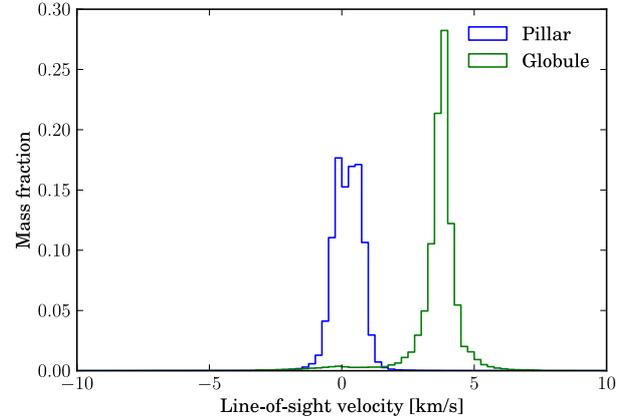}
\caption{\label{histpg} Mass fraction of the gas in the pillar and in
  the globule in function of the velocity. The bulk line-of-sight
  velocity of the pillar is at 0 km/s whereas the bulk velocity of the
  globule is at +4 km/s.} 
\end{figure}

Contrary to the dense clumps in the shell, globules are
  bubbles of gas disconnected from the molecular cloud and surrounding
  by ionized gas. The major difference between the turbulent
simulations presented in this paper and the non-turbulent set-ups
studied in paper I is the appearance of these globules in the highly
turbulent case. As we have already discussed above, globules emerge
because the ram pressure of turbulence dominates the pressure of the
ionized gas. These bubbles of cold dense gas have enough kinetic
energy to penetrate inside the HII region. An interesting consequence
is that the motion of the globules is imposed by the turbulence. They
could have by chance a motion aligned with the direction of expansion
of the HII region but most of the time, they should have a random
motion direction set by the turbulence, and it is indeed the case in
our simulation. The direction of expansion of the HII region is from
top to bottom in Fig. \ref{crho}. Therefore projected on the line of
sight which is perpendicular, the velocity in the expansion direction
is zero. The motion of a pillar is typically aligned with the
direction of expansion of the HII region, as it can be seen on
Fig. \ref{los_60}. The spectra are centered at zero. However it is not
the case for a globule. Figure \ref{los_globule} shows a typical
globule in the simulation at Mach four. The line-of-sight velocity
spectra are systematically shifted at a velocity of +4 km/s which is
of the order of ten times the sound speed of the cold gas. The
  shift is also highlighted in Fig. \ref{histpg}. The globule does
not have a motion aligned with the direction of expansion. In a region
where globules and pillars are present, the velocity of a pillar is
the velocity of expansion of the front projected on the line of sight,
whereas the velocity of a globule will be the signature of the initial
turbulence. The dense clumps that are observed in the shell
  also move with the expansion of the HII region. Therefore the same
  velocity shift between these clumps and the globules nearby can be
  expected. In that picture, pillars and clumps at the edge
  of the HII region are the result of a structure dominated by
the ionization dynamics whereas globules are turbulent dominated. This
is quite different from the radiation driven implosion scenario, in
which there is no reason for the globules to have a velocity different
from pillars. \textbf\\ 
\begin{figure*}[!ht]
\centering
\includegraphics[width=0.32\linewidth]{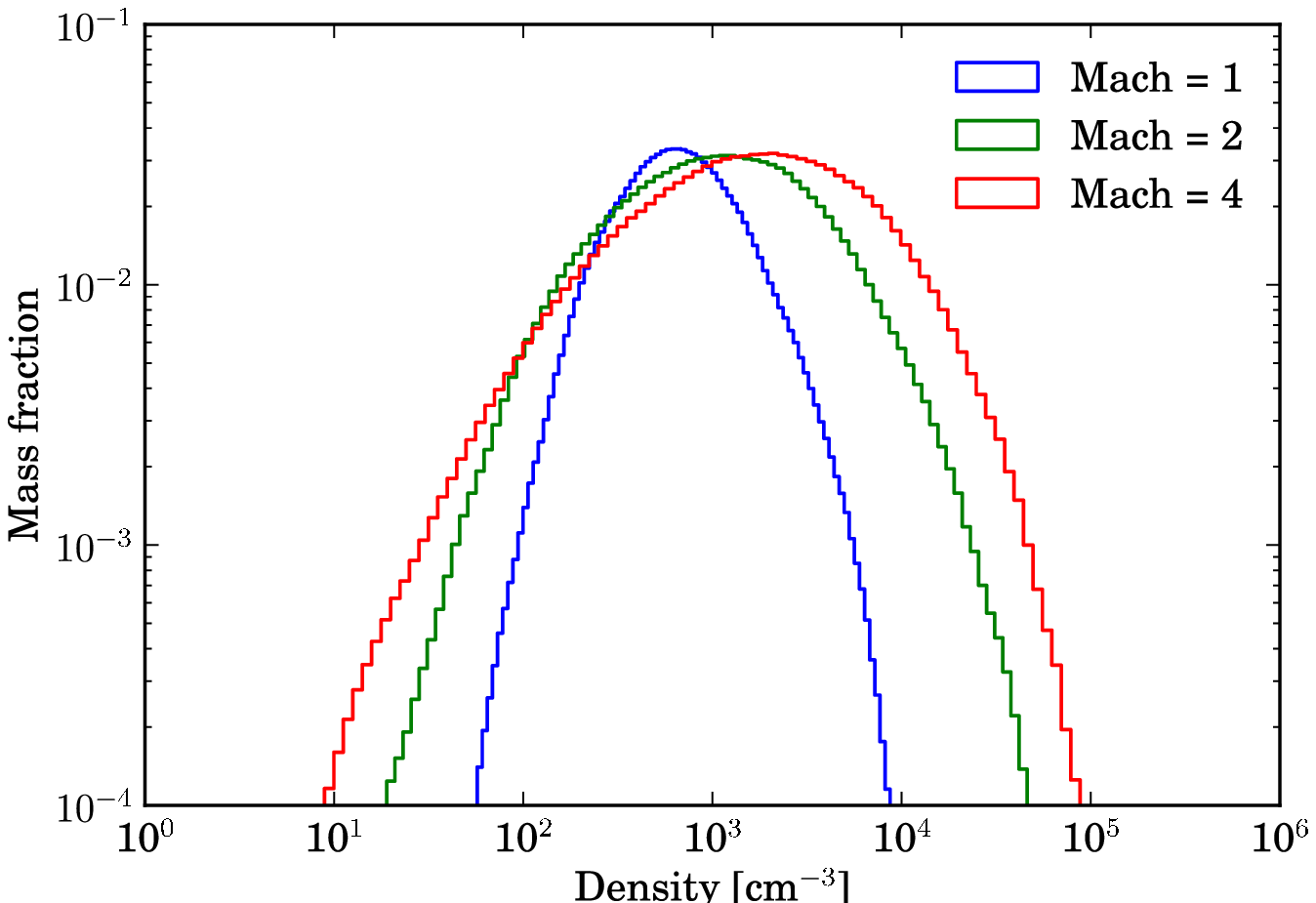}
\includegraphics[width=0.32\linewidth]{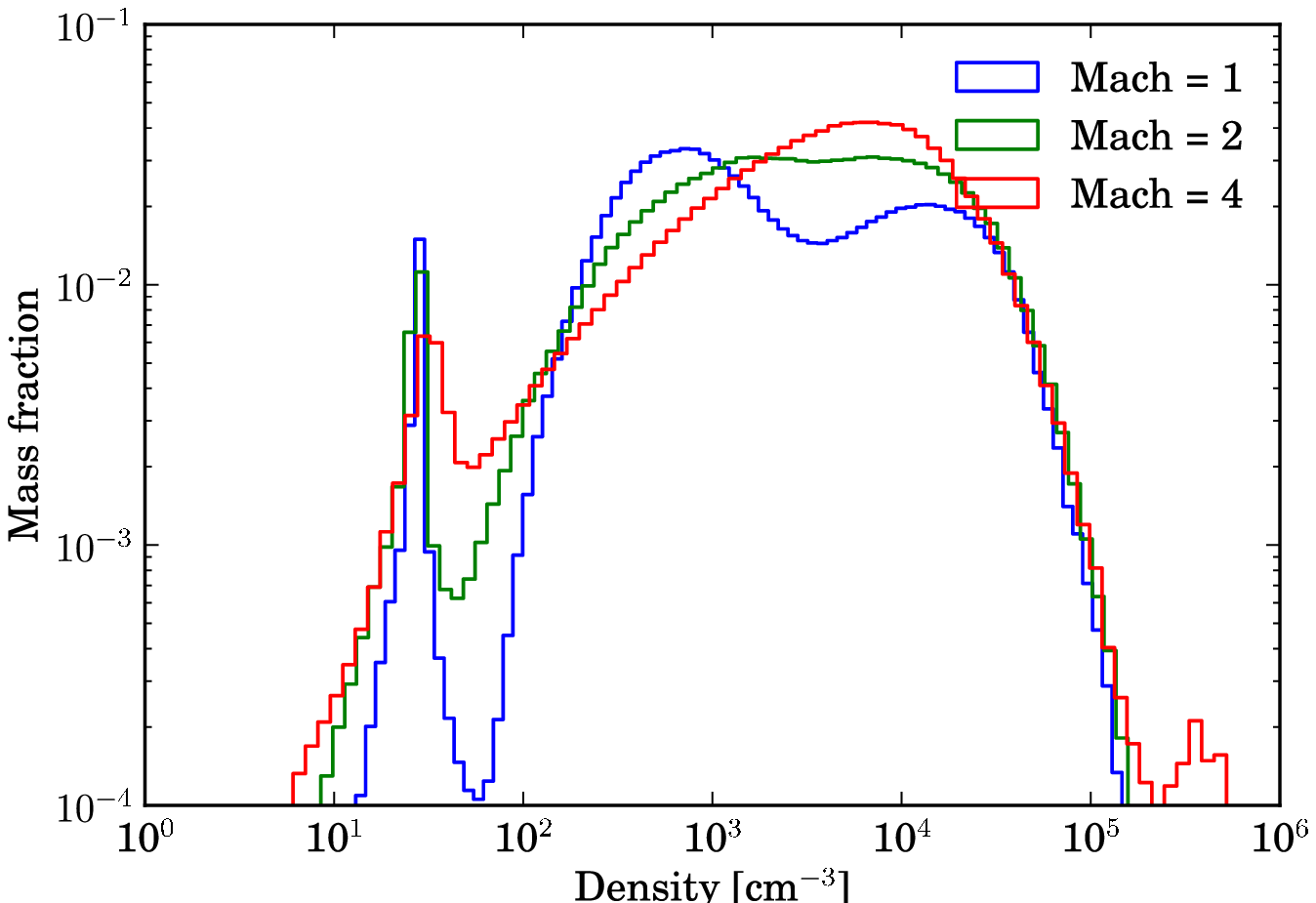}
\includegraphics[width=0.32\linewidth]{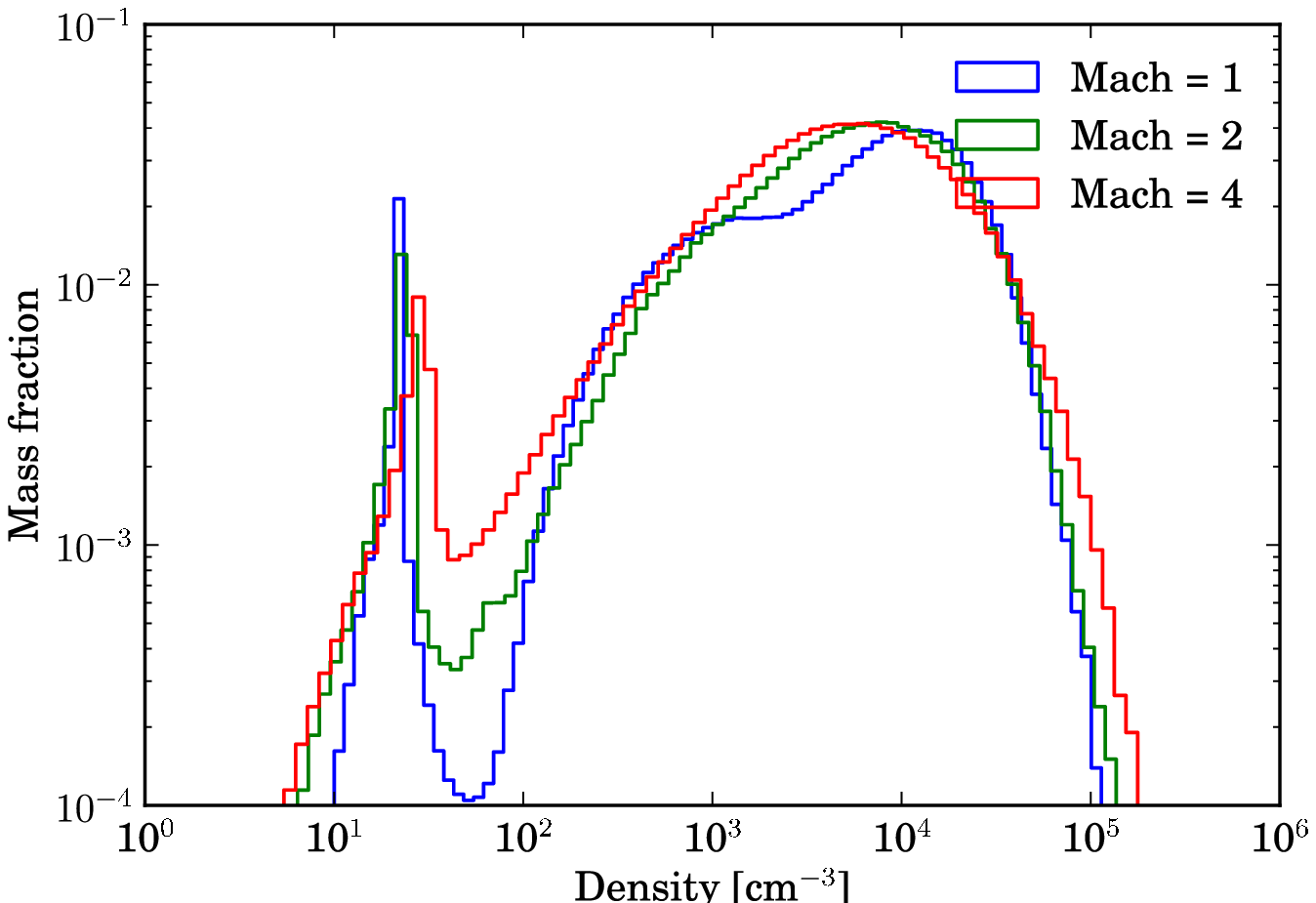}
\caption{\label{pdf} Probability density function of the gas for the
  three simulations. Left: t= 0 ky, middle: t = 500 ky, right: t = 1
  My. The first peak is the ionized gas and therefore is not seen in
  the observations. At t = 500 ky, the Mach-1 simulation presents a
  double peak, the first one corresponds to the unperturbed turbulent
  gas, the second one to the shocked region. In the Mach-2
  simulations, the peak is barely visible because the turbulence is
  higher and because the unperturbed region still present in the box
  becomes small. In the Mach-4 simulation, the turbulence is too high,
  and the second peak is hidden in the peak of the turbulent cold
  region.} 
\end{figure*}
\indent This signature of the turbulent nature of a globule is
supported by recent observations in the Cygnus X region. Thanks to
Herschel Open Time and SOFIA observations \citep{Schneider:2012}, 138
$\mu$m and CO maps are available and contain several pillars and
globules. A pillar and a globule in this region were identified to be
sufficiently close (and the HII region sufficiently big) to consider
they should have the same direction of motion if the motion is only
imposed by the HII region expansion. However the [CII] spectra show a
line-of-sight velocity difference of two kilometers per second, which
is clearly supporting a turbulent scenario for the formation of the
globule against the radiation implosion of an isolated clump. \\ 
\indent The difference of 2 km/s is therefore an indication of the
level of initial turbulence. For a sonic velocity of 0.4 km/s, this
shift indicates typically a motion at Mach 5. In our simulation, the
globule is at Mach 10 whereas the mean Mach number of the simulation
is 4. Therefore the velocity of the globule is an indication of the
turbulence level but not a precise measurement. 

\subsection{Probability density function}

In this section, we investigate the density structure of the gas using
probability density functions (PDFs). Numerical
models
\citep[e.g.][]{Kritsuk:2007gn,Audit:2010jx,BallesterosParedes:2011co}
and observations of 
atomic and molecular gas \citep[e.g.][]{Kainulainen:2009jh} show that PDFs
are log-normal at low densities and can have more complex shapes at
higher densities. We here plot the mass fraction against the density
(Fig. \ref{pdf}) from out simulations in a temporal evolution and obtain that
the PDF shape depends on turbulence and time. At low turbulence (Mach
1) the probability density function is double-peaked because of the
ionization, similarly to the bimodal PDFs caused by the
  thermal instability
  \citep[see][]{SanchezSalcedo:2002fp,Gazol:2005go,Audit:2005df}. The
dense compressed gas is forming a new peak in the distribution. At
high turbulence, the probability density function is shifted toward
higher densities but does not present the double peak. This directly
tell that a double peak is the signature of a region dominated by the
ionization whereas a single peak is the signature that the turbulence
is dominant. This signature was already observed in the Rosette nebula
by \citet{Schneider:2012b}. The PDFs at the edge of the HII regions
present the same double-peaked signature. It suggests that at the
border of the bubble, the turbulence is low compared to the
ionized-gas pressure. \\ 
\indent Based on the PDFs of the Mach 2 simulation, it is clear that
the double peak disappears when there is no more unperturbed gas in
the region on which the PDFs are done. Therefore a good way to look at
the PDFs around a HII region is to do concentric PDFs centered on the
central ionizing cluster. By increasing the radius of the region, we
will first capture the peak of the shocked region, then a double peak
should appear when the unperturbed region is included. Then the first
peak should disappear at some point, when the shocked region becomes
negligible compared to the unperturbed region. This point could
indicates on which scale the central cluster has an impact on the
global PDF of the cloud and therefore on which scale, the initial mass
function is impacted by the ionization. If the turbulence is high, it
is possible that the shocked peak is hidden in the large peak of the
turbulent cloud. In that case, no double peak will appear and this
will be an indication that the turbulence in the region dominates over
the ionized-gas pressure.  

%
%

\section{Conclusions}

Structures at the interface between HII regions and molecular clouds
can be classified mainly in three categories: pillars, globules and
dense clumps. Various scenarii have been investigated in the past to
explain their formation: collect \& collapse, radiation driven
implosion, shadowing effects and turbulence. We presented a new model
to explain the formation of clumps and pillars in paper I based on the
curvature of the dense shell formed by the collect processes. High
curvature leads to pillars, low curvature to instabilities forming
dense clumps and dips in the shell. This model was investigated using
non-turbulent set-ups. In the present paper we show that the same
mechanisms are at work in simulations with a turbulent
cloud. Especially the same spectral observational signatures can be
identified on the formation stages of the pillars.\\ 
\indent Furthermore we have shown that:
\begin{itemize}
\item Because of turbulence, hot ionized gas can get into the shadow
  of cold dense gas, leading to a very long recombination time for the
  gas. This gas can only be treated with a non-equilibrium model for
  ionization, up to 20 \% of the box can get into that state. The
  equilibrium assumption is valid in situations in which the ram
  pressure of turbulence is not dominating the ionized-gas pressure,
  i.e. low turbulence levels. 
\item Globules are formed preferentially when the turbulence in the
  cold gas dominates the ionized-gas pressure. Bubbles of dense gas
  have sufficient kinetic energy to penetrate into the HII regions
  forming the globules. A signature of this scenario is the
  line-of-sight velocity spectrum of the globules which is either
  blue-shifted or red-shifted compared to the spectrum of a pillar
  or a clump at the edge of the HII region. This can be
  directly observed in regions where pillars, clumps and
    globules are present.  
\item Probability density functions are double peaked when the
  turbulent ram pressure is low compared to the ionized-gas
  pressure. They could be used in observations as probes to determine
  the relative importance of the turbulence compared to the
  ionization.  
\end{itemize}

An important step now is to do the statistics of the formation of
dense structures at the edge of HII regions. It will be of great
importance to conclude on the potential negative or positive effects
of radiative feedback on the star-formation rates. Furthermore
comparisons with observations will also tell how realistic the
curvature scenario for the formation of dense clumps and pillars and
the turbulent scenario for the formation of globules are.  

%
%


%
%

\bibliographystyle{aa}
\bibliography{bib_paper.bib}


\end {document}